\documentclass[aps,prb,twocolumn,amsmath,amssymb,nofootinbib,superscriptaddress,floatfix,longbibliography]{revtex4-2}

\usepackage{xcolor}
\usepackage{booktabs}
\usepackage{amsmath}
\usepackage{amssymb}
\usepackage{amsthm}
\usepackage{mathtools}
\usepackage{mathdots}
\usepackage{amsfonts}
\usepackage{bbm}
\usepackage{xr}
\usepackage{bbold}
\usepackage{epsfig,color,dsfont,upgreek,physics}
\usepackage{mathrsfs}
\usepackage{multirow}
\usepackage[makeroom]{cancel}
\usepackage{mathtools}

\usepackage{graphicx}
\usepackage{dcolumn} 
\usepackage{bm} 
\usepackage{float} 
\usepackage[driverfallback=dvipdfm]{hyperref} 
\usepackage{ulem}   
\hypersetup{
    colorlinks=true,
    linkcolor=blue,
    citecolor=blue,
    urlcolor=blue
}

\allowdisplaybreaks

\newcommand{\bs}[1]{{\boldsymbol{#1}}}

\begin{document}

 \title{Electromagnetic response and emergent topological orders in transition metal dichalcogenide 
\texorpdfstring{MoTe$_2$}{} bilayers
}

\author{Tianhong Lu}
\affiliation
{
Department  of  Physics,  Emory  University,  400 Dowman Drive, Atlanta,  GA  30322,  USA
}

\author{Yi-Ming Wu}
\affiliation
{
Stanford Institute for Theoretical Physics, Stanford University, Stanford, California 94305, USA
}

\author{Luiz H. Santos}
\affiliation
{
Department  of  Physics,  Emory  University,  400 Dowman Drive, Atlanta,  GA  30322,  USA
}

\begin{abstract}
Twisted bilayer transition metal dichalcogenides, such as MoTe$_2$, provide a versatile platform for exploring correlated topological phases. This work investigates the interplay of perpendicular magnetic and electric fields in tuning the electronic structure and emergent topological orders of twisted bilayer MoTe$_2$ (t-MoTe$_2$) across two distinct regimes: a low-twist-angle phase ($\theta\approx2.1^\circ$) hosting multiple Chern bands of identical Chern numbers per valley, and a higher-angle phase ($\theta\approx 3.89^\circ$) featuring Haldane-like bands with opposite Chern numbers. Using a continuum model incorporating moir\'e potentials up to second harmonics, we compute the Hofstadter fractal spectra under applied fields, revealing Landau fan structures and magnetic-flux-dependent band topology. 
These fractal spectra are useful in studying emergent topological orders in terms of the composite fermion picture, where the statistical Chern-Simons flux is approximated as a uniform gauge field.
We demonstrate that the system hosts both Jain-sequence fractional Chern insulators (FCIs) and non-Jain ``fractal FCIs" with higher Chern numbers. The electric field suppresses composite fermion gaps and induces topological quantum phase transitions. Furthermore, our analysis extends to valley-contrasting flux attachment, proposing pathways to describe fractional quantum spin Hall states. 
\end{abstract}

\date{\today}

\maketitle

\section{Introduction}
The method of stacking and twisting two-dimensional van der Waals materials opens up new and promising directions for creating topological electronic bands, offering platforms for novel phases of matter \cite{andrei2021marvels}. In particular, twisted bilayer MoTe$_2$ (t-MoTe$_2$), a member of the transition metal dichalcogenide (TMD) family, stands out as a remarkably rich class of moir\'e materials that has garnered significant attention. Due to its strong spin-orbit coupling and broken inversion symmetry, the electronic states of single-layer MoTe$_2$ feature a spin-valley locking in the valence bands and a valley-contrasting Berry curvature near each valley center \cite{xiao2012coupled}.
Compared to graphene, the spin-valley locking in hole-doped TMDs simplifies the electronic degrees of freedom while keeping the essential topology intact, making them advantageous platforms for investigating topological moir\'e flat bands. 
Indeed, when two layers of MoTe$_2$ are stacked and rotated by a small angle $\theta$, these unique features
give rise to a set of tunable, nearly flat topological moir\'e bands \cite{fengcheng2019topological, wang2024fractional, wang2023topology, Ahn2024non-abelian, zhang2024polarization, Xu2025multiple, wang2025higher}.

The moir\'e band structure of t-MoTe$_2$ and its topological properties are highly sensitive to the twist angle. 
For $2.0^{\circ} \lesssim \theta \lesssim 2.7^{\circ}$, 
the first few hole-doped moir\'e bands constructed from one valley all have the same Chern number $C=1$, while those constructed from the other valley have $C=-1$ instead, as required by time-reversal symmetry \cite{zhang2024polarization, Xu2025multiple, wang2025higher}. 
As the twist angle increases, several topological transitions occur. In the range $2.8^{\circ} \lesssim \theta \lesssim 3.9^{\circ}$, the topmost pair of bands per valley exhibit Chern numbers $C=\pm 1$, resembling a pair of Haldane bands \cite{Haldane1988}.

The remarkable tunability of band topology through the twist angle profoundly influences the array of emergent phenomena in hole-doped t-MoTe$_2$. For $\theta \sim 3.7^{\circ} - 3.9^{\circ}$, partial filling of the Haldane-like bands gives rise to fractional Chern insulators \cite{cai2023signatures, park2023observation, zeng2023thermodynamic, xu2023observation, park2025ferromagnetism, park2025observation, xu2025signatures, redekop2024direct} even in the absence of an external magnetic field. Furthermore, at $\theta \approx 2.1^{\circ}$, where multiple valley Chern bands with the same Chern number emerge, strong signatures of integer quantum spin Hall states appear at integer band fillings, accompanied by incipient fractional quantum spin Hall states near half-filling of the second moir\'e band \cite{Kang2024evidence, kang2025time}. 

These findings underscore the potential of t-MoTe$_2$ to host a diverse range of chiral and non-chiral topological phases, many of which remain largely unexplored. They also emphasize the critical role of understanding the landscape of topological bands and their dependence on experimentally tunable parameters. Beyond the twist angle, key tuning factors include perpendicular electric and magnetic fields, which can be imprinted on the moir\'e sample in a dual-gate setup.

In this work, we characterize the low-energy hole-doped moir\'e bands of t-MoTe$_2$ as a function of perpendicular electric ($E$) and magnetic ($B$) fields. Our analysis focuses on two distinct twist angle regimes: \textbf{Phase I} ($\theta \approx 2.1^\circ$), which hosts multiple
 Chern bands of the same Chern number per valley as shown in Figure \ref{fig: moire bands} (a), and \textbf{Phase II} ($\theta \approx 3.9^\circ$), where a pair of Haldane-like bands with opposite Chern numbers per valley emerges, as shown in Figure \ref{fig: moire bands}(b).  
We employ a continuum model that describes the valley states of bilayer TMD valence bands \cite{fengcheng2019topological}, incorporating moir\'e potentials on the top and bottom layers along with tunneling terms. For a more accurate calculation, we also include both the first and second harmonics \cite{wang2024fractional, wang2023topology, Ahn2024non-abelian, zhang2024polarization, Xu2025multiple, wang2025higher}. This approach allows for a direct implementation of electric field effects - modeled as a potential energy difference between top and bottom layers, tunable in the dual-gate setting - and magnetic field effects via gauge-invariant minimal substitution.

The energy spectrum of electrons moving in a two-dimensional periodic lattice and in the presence of a perpendicular magnetic field exhibits a Hofstadter fractal structure\cite{Hofstadter76}. While the observation of such fractal spectra in atomic-scale lattices requires impractically strong magnetic fields, moir\'e
superlattices offer an ideal platform to investigate fractal electronic states
owing to the large nanometer-scale moir\'e period where the magnetic flux per moir\'e unit cell can reach values comparable to a flux quantum $\phi_0 = h/e$ for
experimentally accessible fields. While significant theoretical \cite{Bistritzer2011moire,hejazi2019landau,wang_classification_2020, Herzog-Arbeitman-Hofstadter2020} and experimental \cite{Dean13, Ponomarenko13, Hunt13, Forsythe18, Spanton18, das2021observation,das2021symmetry,BalentsEfetovYoung20,stepanov2021competing,Saito21,wu2021chern} efforts have been devoted the characterization of fractal electronic states in graphene-based superlattices, the character of such fractal bands in twisted bilayer MoTe$_2$ remains much less explored. 

A central goal of this work is to characterize the intricate landscape of fractal energy bands in twisted MoTe\(_2\), tunable by external electric and magnetic fields and experimentally accessible in Phases I and II. Our analysis captures the interplay between these fields while incorporating moir\'e potentials up to second harmonic contributions. This approach extends and complements previous studies: Ref.~\cite{wang2024phase} focused on the Hofstadter spectrum near a \(3.9^\circ\) twist angle, using a first harmonic approximation for the moir\'e potential, while Ref.~\cite{kolar2024hofstadter} explored the \(2.1^\circ\) regime including second harmonics, but without accounting for a displacement field. By unifying and advancing these perspectives, our study offers a comprehensive framework for uncovering emergent topological bands and quantum phase transitions in t-MoTe\(_2\), paving the way for future investigations into novel correlated states and competing electronic orders.

Furthermore, the study of gauge fields coupled to moir\'e bands also offers a powerful framework for characterizing emergent topological orders when the bands of t-MoTe$_2$ are partially filled. In particular, the observation of zero-field fractional Chern insulators (FCIs) in t-MoTe$_2$ at $\theta \sim 3.7^\circ - 3.9^\circ$ \cite{cai2023signatures, park2023observation, zeng2023thermodynamic, xu2023observation, park2025ferromagnetism, park2025observation, xu2025signatures, redekop2024direct} at hole filling fractions $\nu_{h} = 2/3, 3/5, 4/7, 5/9$ mirroring the Jain sequence of Abelian fractional quantum Hall states in an external magnetic field, provides compelling evidence that the relevant degrees of freedom of the system can described by composite fermions characterized by the binding of two statistical Chern-Simons flux quanta per particle \cite{lu_santos_2024fractional}.

Motivated by this phenomenological picture, we carry out a comprehensive analysis of composite fermion states emerging from the Hofstadter spectrum induced by a mean-field uniform Chern-Simons flux in both Phases I and II. In this framework, the Hofstadter spectrum at finite external magnetic field can be directly transposed to the composite fermion picture, revealing a rich sequence of valley-polarized fractional Chern insulator (FCI) states. We map out the resulting incompressible phases as a function of filling fraction per valley and extract the fractional Hall conductivity from the topological properties (Chern numbers) of the filled composite fermion bands. A key novelty, in contrast to the conventional fractional quantum Hall effect (FQHE) in continuum Landau levels, is the emergence of composite fermion bands within a fractal Hofstadter spectrum, possessing Chern numbers distinct from those of Landau levels. These give rise to fractal FCIs \cite{lu_santos_2024fractional}
where the Hall conductivity and filling fraction deviate from the standard Jain hierarchy~\cite{jain1989composite,lopez1991fractional} due to the significant influence of the moir\'e superlattice. It is worth noting that such emergent topological orders have also been numerically reported in \cite{chen2025fractional} recently. Although these fractal FCIs tend to exhibit smaller composite fermion gaps, their experimental realization would mark an important advance in the study of fractionalized topological phases.

Our study also highlights the electric field as a powerful tuning parameter for accessing distinct topological phases in t-MoTe$_2$. Specifically, we find that the electric field suppresses the composite fermion gaps associated with the Jain sequence. More strikingly, it can induce topological quantum phase transitions within fractal FCIs,
such as those at filling fractions $\nu_h = 4/5$, $2/9$, and $1/5$, which
could be within experimental reach.

We further extend the composite fermion approach to describe topological orders with time-reversal symmetry through a valley-contrasting flux attachment mechanism, leading to a novel series of Abelian fractional quantum spin Hall states~\cite{bernevig2006quantum,levin2009fractional,santos2011time,neupert2011fractional}. These states emerge as valley FCIs that transform under time-reversal symmetry. Recent work has highlighted the potential of moir\'e flatbands as a platform for realizing such fractional quantum spin Hall phases~\cite{wu2024time-reversal,abouelkomsan2025non, jian2024minimal, may2024theory, sodemann2024halperin, chou2024composite, kwan2024could}, 
which remains under active experimental investigation \cite{Kang2024evidence,kang2025time}. Our results thus lay the groundwork for analyzing non-chiral topological orders in moir\'e systems, an area where the underlying mechanisms remain largely unexplored.

The paper is organized as follows. In Sec.\ref{sec:continuum model} we review the continuum model and the band structure in the absence of external fields. In Sec.\ref{sec:couplingtoB} we present our method and convention of coupling t-MoTe$_2$ to magnetic and electric fields, and the results are presented in Sec.\ref{sec:zeroEfield} and \ref{sub:moir_e_hofstadter_bands_at_non_zero_displacement_field}. In Sec.\ref{sec:CF states} we apply the Hofstadter spectrum to composite fermions by assuming a uniform Chern-Simons flux attachment, and identify different FCI states and how they response to displacement field. In Sec.\ref{sec:discussion} we summarize the results and outline some future directions. Some technical details and supplementary results are presented in Appendcies.

\section{Overview of the Band structure of \texorpdfstring{t-MoTe$_2$}{}  without field effects}
\label{sec:continuum model}

\subsection{The continuum model}
\label{sec:The continuum model}

In this section, we review the continuum model of twisted bilayer MoTe$_2$. 
The low-energy states of AA-stacking t-MoTe$_2$ is formed by spin-polarized K and K' valleys, which are related by time-reversal symmetry. The twist angle $\theta$ leads to a relative shift of the K (K') points in the two layers by $k_\theta = \frac{4\pi\theta}{3a_0}$ (where $a_0$ is the lattice constant of monolayer Mote$_2$), 
\begin{figure}[!htb]
    \centering
    \includegraphics[width=1\linewidth]{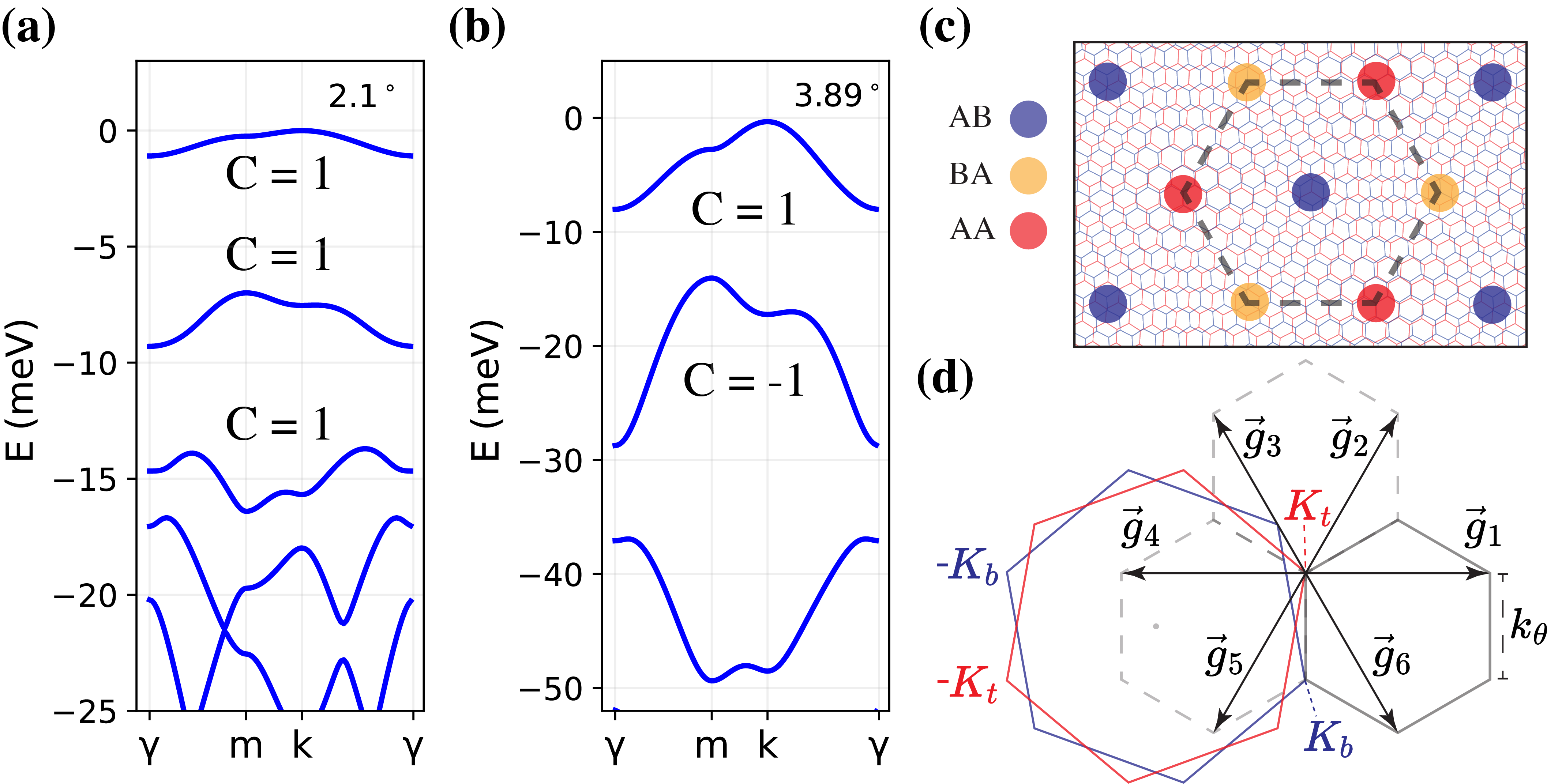}
    \caption{(a,b) The moir\'e bands (K valley) at 2.1 and 3.89 degrees.
    (c) The moir\'e lattice, where the AA, AB, BA centers are denoted.
    (d) The moir\'e Brillouin zone, where the reciprocal vectors are highlighted.}
    \label{fig: moire bands}
\end{figure}
into the corners of the moir\'e BZ,
i.e., $\boldsymbol{k}_{b/t} = k_\theta(-\frac{\sqrt{3}}{2},\mp\frac{1}{2})$ for the bottom/top layers, as shown in Fig. \ref{fig: moire bands}(d).
The low-energy Hamiltonian for $(K, \uparrow)$ states is described by the by the single particle Hamiltonian\cite{fengcheng2019topological}
\begin{equation}
\label{eq: polarized continuum model K valley}
\mathcal{H}_{K,\uparrow}^0(\boldsymbol{p},\bs{r}) = \begin{pmatrix}
\frac{-\hbar^2(\bs{p}-\bs{k}_{b})^2}{2m^{*}} + V_{b}(\bs{r})
&\,
T(\bs{r})
\\
T^{\dagger}(\bs{r})
&\,
\frac{-\hbar^2(\bs{p}-\bs{k}_{t})^2}{2m^{*}} + V_{t}(\bs{r})
\,
\end{pmatrix}
\,,
\end{equation}
where $\bs{p} = -i\hbar\bs{\nabla}$ is the momentum operator, $m^*$ is an effective mass, $V_{b}(\bs{r})$ and $V_{t}(\bs{r})$ denote, respectively, the intra-layer moir\'e potentials in the bottom and top layers, and $T(\bs{r})$ captures the inter-layer tunneling amplitude. These two terms can be expressed in terms of the first and second harmonics,
$V(\bs{r}) = V^{(1)}(\bs{r}) + V^{(2)}(\bs{r})$ and $T(\bs{r}) = T^{(1)}(\bs{r}) + T^{(2)}(\bs{r}),$ 
respecting the periodicity of the emergent moir\'e superlattice (Fig. \ref{fig: moire bands}(c)), {characterized by the moir\'e lattice vectors $\boldsymbol{a}_1=\frac{a_0}{\theta}(\sqrt{3}/2,-1/2),\boldsymbol{a}_2=\frac{a_0}{\theta}(0,1)$}.

The first and second harmonics are, respectively,
\begin{equation}\label{1stHarmonics}
    \begin{aligned}
V^{(1)}_b(\mathbf{r}) & =V e^{i \psi}\left(e^{i \mathbf{g}_1 \cdot \mathbf{r}}+e^{i \mathbf{g}_3 \cdot \mathbf{r}}+e^{i \mathbf{g}_5 \cdot \mathbf{r}}\right)+\text { h.c. }\,, \\
V^{(1)}_{t}(\mathbf{r}) & =V^{(1)}_{b}(-\mathbf{r})\,, \\
T^{(1)}(\mathbf{r}) & =w\left(1+e^{-i \mathbf{g}_2 \cdot \mathbf{r}}+e^{-i \mathbf{g}_3 \cdot \mathbf{r}}\right)
\,,
\end{aligned}
\end{equation}
\begin{equation}\label{2ndHarmonics}
\begin{split}
V_{b/t}^{(2)}(\mathbf{r})& = V_2\left[e^{i\left(\mathbf{g}_1+\mathbf{g}_2\right) \cdot \mathbf{r}}+e^{i\left(\mathbf{g}_3+\mathbf{g}_4\right) \cdot \mathbf{r}}+e^{i\left(\mathbf{g}_5+\mathbf{g}_6\right) \cdot \mathbf{r}}\right]\,,\\
T^{(2)}(\mathbf{r})&=w_2\left[e^{i \mathbf{g}_1 \cdot \mathbf{r}}+e^{i \mathbf{g}_4 \cdot \mathbf{r}}+e^{i\left(\mathbf{g}_3+\mathbf{g}_2\right) \cdot \mathbf{r}}\right]\,,
\end{split}
\end{equation}
where $\boldsymbol{g}_i=R^{i-1}_{\pi/3}(g,0)$, with $g=\sqrt{3}k_{\theta}$, are reciprocal lattice vectors denoted in Fig.\,\ref{fig: moire bands}(d).

The parameters in Eqs.\,\eqref{1stHarmonics} and \eqref{2ndHarmonics} can be extracted by fitting to density functional theory (DFT) bands~\cite{zhang2024polarization,wang2024fractional}. In this work, we adopt the values ($a_0$, $m^*$, $V$, $\psi$, $w$, $V_2$, $w_2$) = ($0.3472\mathrm{~nm}$\cite{liu2013three,agarwal1972measurement}, $0.62 \mathrm{~m}_{\mathrm{e}}$, $20.51 \mathrm{meV}$, $-61.49^{\circ}$,$-7.01 \mathrm{meV}$, $-9.08 \mathrm{meV}$, $11.08\mathrm{meV}$) at $\theta=2.1^\circ$ and ($0.355\mathrm{~nm}$\cite{wang2023topology}, $0.62 \mathrm{~m}_{\mathrm{e}}$, $9.45 \mathrm{meV}$, $-85.23^{\circ}$,$-12.2 \mathrm{meV}$, $24.99 \mathrm{meV}$, $13.12\mathrm{meV}$)
at $\theta=3.89^\circ$
from Ref.\cite{Ahn2024non-abelian}. With these parameters, one can diagonalize the Hamiltonian in a plane wave basis, leading to the moir\'e band structures as shown in Fig.\ref{fig: moire bands}(a) and (b).

\section{Band structure of 
\texorpdfstring{t-MoTe$_2$}{} 
in external \texorpdfstring{$E$ and $B$}{} fields}
\label{sec:bands with E and B fields}

We investigate the effects of external perpendicular electric and magnetic fields coupled to spin-polarized valley states. These effects are incorporated in the low-energy Hamiltonian \eqref{eq: polarized continuum model K valley} via
\begin{equation}
\label{eq: continuum model with B and E fields}
\mathcal{H}_{K,\uparrow}(\boldsymbol{p},\bs{r}) = 
\mathcal{H}_{K,\uparrow}^0(\boldsymbol{p}+e\,\boldsymbol{A}(\bs{r}),\bs{r})+\frac{u_d}{2}\sigma_z
\,,
\end{equation}
where $u_{d}$ describes the electrostatic potential difference between the bottom and top layers, and 
the magnetic field $\mathbf{B}=\nabla \times \mathbf{A}=- B \hat{\mathbf{z}}$ enters via minimal coupling $\bs{p} \rightarrow \bs{p} + e\bs{A}(\bs{r})$, where $e$ is the hole charge and $\bs{A}(\bs{r})$ is the vector potential. 
Although our focus will be on the orbital effects from the magnetic field, we note that the zeeman effect, if large, will lead to valley polarization of the moir\'e bands, and we will comment on this below.

The large moir\'e unit cell can support magnetic flux per unit cell comparable to a flux quantum $\phi_0$, giving rise to moir\'e Hofstadter states. In the following, we discuss how this band structure of a single valley breaks into fractal Hofstadter bands when the unit cell supports a magnetic flux $\phi = B\,A_{uc} = \frac{p}{q}\,\phi_0$, for coprime integers $p$ and $q$, where the area of the moir\'e unit cell is
$A_{uc} = |\bs{a}_1 \times \bs{a}_2| = \frac{4\pi^2}{|\vec{g}_1\times\vec{g_2}|}=\frac{8\pi^2}{\sqrt{3}|\vec{g}_1|^2}
$. To this end, we shall diagonalize the Hamiltonian in Eq.\eqref{eq: polarized continuum model K valley} using Landau level basis instead of plane wave basis.

\subsection{Coupling to magnetic fields: Hofstadter bands}
\label{sec:couplingtoB}

In the absence of the moir\'e potential, the single particle eigen states of the Hamiltonian \eqref{eq: continuum model with B and E fields} are given by decoupled Landau levels in each layer, spaced by the cyclotron energy $\hbar\omega_c$, where $\omega_c = \frac{eB}{m^{*}}$ is the cyclotron frequency. We express the position $\bs{r} = \bs{R} + \bs{\eta}$ in terms of the guiding center coordinate $\bs{R}$ and the cyclotron coordinate $\boldsymbol{\eta}=\ell_B^2 \hat{\mathbf{z}} \times \boldsymbol{\Pi}$ (where $\bs{\Pi} = \frac{1}{\hbar}(\bs{p} + e\,\bs{A})$), which obey commutation relations
\begin{equation}
    \left[\eta_x, \eta_y\right]=i \ell_B^2,\quad
    \left[R_x, R_y\right]=-i \ell_B^2\,, \quad
    [\boldsymbol{R}\,,\boldsymbol{\eta}]=0\,,\label{eq:Reta}
\end{equation}
where $\ell_B = \sqrt{\hbar/eB}$ is the magnetic length. 
The particle density
per LL is $\rho_{LL} = \frac{1}{2\pi\ell^{2}_{B}}$, and each LL supports $N_{LL} = \rho_{LL}\,A_{\textrm{system}}$ states, where $A_{\textrm{system}}$ is the system area.

The moir\'e potential $V_{b/t}(\bs{r})$ and layer tunneling $T(\bs{r})$ lead to mixing and broadening of these Landau levels. To account for these effects, {we first note that both $V_{b/t}(\bs{r})$ and $T(\bs{r})$ are expressed as a sum of operators of a particular form $V_{m_1,m_2}=e^{i (m_1 \boldsymbol{g}_1+m_2\boldsymbol{g}_2) \cdot \bs{r}}$ with $m_1$ and $m_2$ being two independent integers. But according to Eq.\eqref{eq:Reta} we can also write  }
\begin{equation}
\label{AmnBmn}
\begin{split}
    V_{m_1,m_2}
    &\,=
    e^{i (m_1 \boldsymbol{g}_1+ m_2 \boldsymbol{g}_2) \cdot \bs{\eta}}\,e^{i (m_1 \boldsymbol{g}_1+ m_2 \boldsymbol{g}_2) \cdot \bs{R}}
    \\
    &\,\equiv \mathcal{A}_{m_1,m_2}\mathcal{B}_{m_1,m_2}\,.
\end{split}
\end{equation}
In the following we express this operator
in the Landau level basis, for reciprocal lattice vectors $\bs{g} = m_1 \bs{g}_1 + m_2 \bs{g}_2$.

{The two reciprocal lattice vector $\bm{g}_1$ and $\bm{g}_2$ are not orthogonal. Thus, instead of working with $R_x$ and $R_y$ \cite{Bistritzer2011moire,hejazi2019landau} as discussed in Eq.\eqref{eq:Reta}, it is more convenient to use}
a natural projection of $\boldsymbol{R}$ to the reciprocal vectors $\boldsymbol{g}_1$ and $\boldsymbol{g}_2$ as $R_1 = \hat{\boldsymbol{g}}_1\cdot\boldsymbol{R}\,, R_2 = \hat{\boldsymbol{g}}_2\cdot\boldsymbol{R}\,,$
which satisfies
\begin{equation}
    [R_1, R_2] = -i\ell_B^{2} (\hat{\bs{g}}_1 \times \hat{\bs{g}}_2)\cdot\hat{z} = -i\frac{\sqrt{3}}{2}\ell_B^2\,,
\end{equation}
and
\begin{equation}
e^{-i g\,R_1}\,R_2 \,e^{i g\,R_1}
= R_2 - \frac{\sqrt{3}}{2}g\,\ell_B^{2}\,.
\end{equation}

We choose a specific gauge such that the Landau level are denoted by $|n,y\rangle$, where $n = 0, 1, 2,...$ is the Landau level index and $y$ labels the guiding center coordinate $R_2$. It follows that
\begin{equation}
\label{eq:action of R1 and R2 on LL basis}
\begin{split}
&\,e^{igR_2}|n,y\rangle = e^{igy}|n,y\rangle,\quad 
e^{igR_1}|n,y\rangle = |n,y-\Delta\rangle\,,
\\
&\,
\Delta = \frac{\sqrt{3}}{2}g\ell_B^2\,, 
\quad  g\Delta = \frac{2\pi q}{p}
\,.
\end{split}
\end{equation}
Eq.\eqref{eq:action of R1 and R2 on LL basis}
implies a $p$-periodicity under which translation invariance is regained as $y \rightarrow y + p\,\Delta$, leading effectively to the folding of each LL into $p$ subbands. 
Note that this is consistent with magnetic translation symmetry. 
In fact, for a magnetic flux $\phi = \frac{p}{q}\phi_0$ per moir\'e unit cell of area ${A}_{uc}$, the magnetic unit cell is formed by enlarging the moir\'e unit cell by a factor of $q$, which leads to the number of states in the magnetic unit cell to be
$\frac{q\,A_{uc}}{2\pi\ell^{2}_{B}}
= q\,\frac{\phi}{\phi_0} = p$, so that each LL gives rise to $p$ ``subbands", each one supporting
\begin{equation}
\label{eq: N sub}
N_{sub} = \frac{N_{LL}}{p} = \frac{N_{uc}}{q} 
\,
\end{equation}
states, where $N_{uc}$ is total number of unit cells.

Thus, upon expressing the guiding center coordinate  as $y = y_0 + (u\,p+j)\Delta$, where $0 \leq y_0 \leq \Delta$,
$u \in \mathbb{Z}$ and $j = 0, 1, ..., p-1$, (with $j \equiv j ~\textrm{mod}$(p)) the action of the guiding center part $\mathcal{B}_{m_1,m_2}$ in Eq.\eqref{AmnBmn} reads
\begin{equation}
\label{Bmn}
\begin{split}
&\mathcal{B}_{m_1,m_2}|n,y_0+(up+j)\Delta\rangle\\ &= 
    e^{-im_1 m_2\frac{\pi q }{p}}e^{i m_2 g(y_0+(up+j)\Delta)}|n,y_0+(up+j-m_1)\Delta\rangle\\
    &=e^{i m_2 g y_0}e^{i\frac{2\pi q}{p}m_2(j-\frac{m_1}{2})}|n,y_0+(u p+j-m_1)\Delta\rangle
    \,.
\end{split}
\end{equation}

Furthermore, using the LL creation and annihilation operators
\begin{equation}
    a=\frac{\eta_x+i \eta_y}{\sqrt{2 \ell_B^{ 2}}}, \quad a^{\dagger}=\frac{\eta_x-i \eta_y}{\sqrt{2 \ell_B^2}},\quad  \left[a, a^{\dagger}\right]=1\,,
\end{equation}
the matrix elements of $\mathcal{A}_{m_1, m_2}$ can be computed from the expression \cite{murthy2003hamiltonian}
\begin{equation}
\label{Laguerre}
\begin{split}
&\,    
\left\langle n_2, y_2\right| e^{\pm i \bs{q} \cdot \bs{\eta}}\left|n_1, y_1\right\rangle
\\
&\,
=
\delta_{y_1, y_2}
\,
\sqrt{\frac{n_{2}!}{n_{1}!}} e^{-x / 2}
(\frac{\pm i q_{-} \ell_B}{\sqrt{2} })^{n_1-n_2} 
L_{n_2}^{n_1-n_2}(x)\,,
\end{split}
\end{equation}
for $n_1 \ge n_2$, where $q_- = q_x -iq_y$, $x = |q_-|^2\ell_B^2/2$ and $L_{n_2}^{n_1-n_2}(x)$ is the Laguerre polynomial (See Appendix A).

Introducing the Fourier transform,
\begin{equation}
    \left|\lambda, n, y_0, j, k_2 \right\rangle=\frac{1}{\sqrt{N}} \sum_u e^{i k_2 (u p+j) \Delta}\left|\lambda, n, y_0+(u p+j) \Delta\right\rangle
\end{equation}
where $\lambda =\{ b,t \}$ denotes the layer index, and $k_2$ is a momentum where $ 0 \leq k_2 \leq \frac{2\pi}{p\,\Delta}$,
we can recast the Hamiltonian Eq.\eqref{eq: polarized continuum model K valley} in this basis
\begin{equation}
\begin{split}
&\,    
\mathcal{H}_{\lambda_1, n_1, y_{0}, j_1, k_2; \lambda_2, n_2, y_{0}, j_2, k_2}   
\\
&\,=
\langle \lambda_1, n_1, y_{0}, j_1, k_2| 
\mathcal{H}
|\lambda_2, n_2, y_{0}, j_2, k_2
\rangle
\,.
\end{split}
\end{equation}
By defining an effective momentum $k_1 \equiv y_0 / \ell_B^2$, with $0 \leq k_1 < \Delta / \ell_B^2$, the single-particle Hamiltonian can be written in Bloch form as $\mathcal{H}_{\lambda_1, n_1, j_1; \lambda_2, n_2, j_2}(k_1, k_2)$, where $(k_1, k_2)$ lie within the first Brillouin zone defined by the Bloch momenta. The dimension of the Bloch Hamiltonian matrix is $2 \times p \times N^{\textrm{max}}_{LL}$, where $N^{\textrm{max}}_{LL}$ denotes the maximum number of LL's in our simulations, which is of the order of {$N^{\textrm{max}}_{LL} \sim$} 200.

\subsection{Moir\'e Hofstadter bands at zero displacement field}
\label{sec:zeroEfield}

\begin{figure}[!htb]
    \centering
    \includegraphics[width=0.9\linewidth]{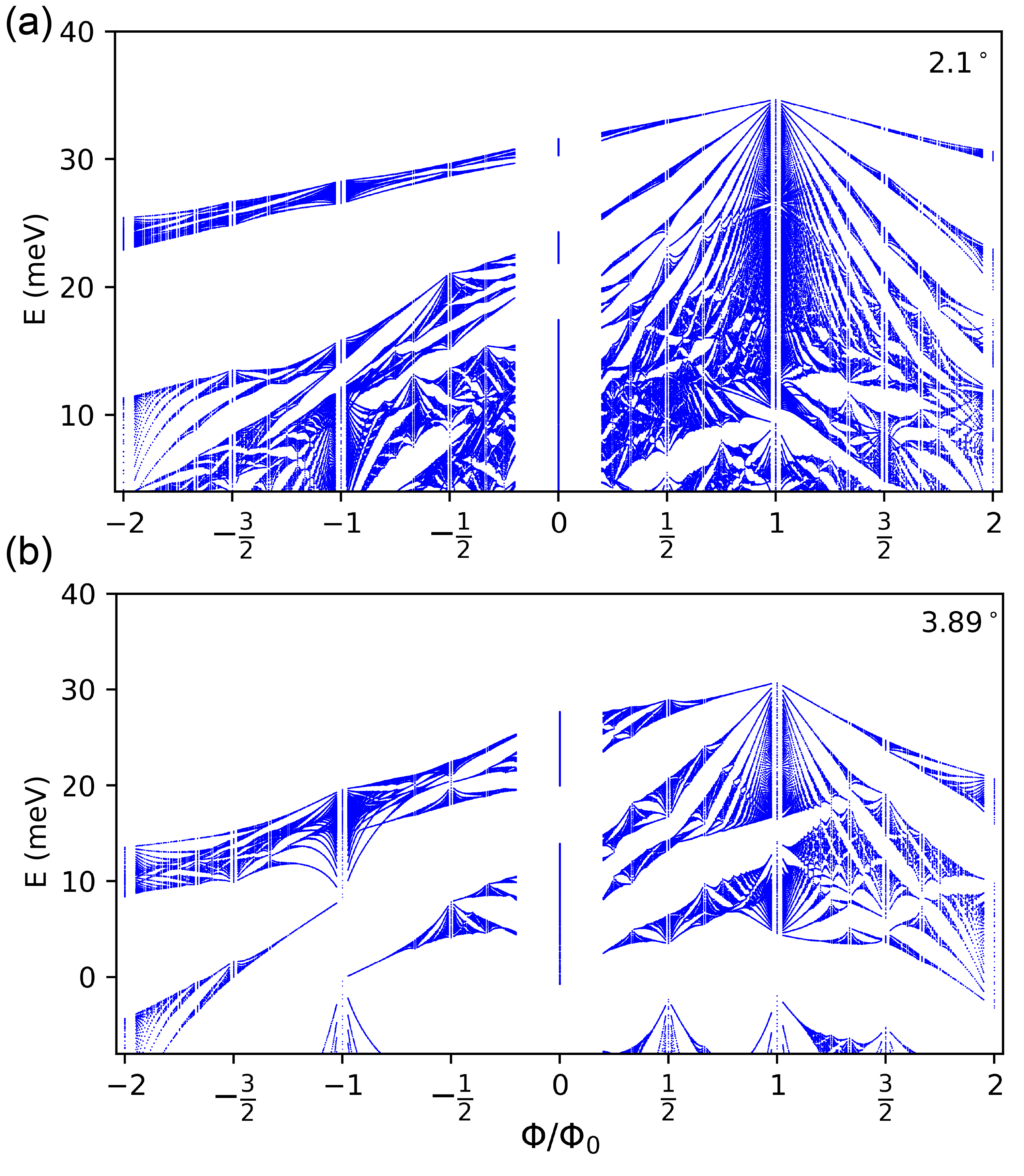}
    \caption{The moir\'e Hofstadter spectra at zero displacement field for (a) $\theta \approx 2.1^{\circ}$ and (b) $\theta\approx3.89^{\circ}$. A Landau fan structure emerges when one flux quantum pierces each moir\'e unit cell, with the Landau level splitting increasing as the flux deviates from $\phi_0$.}
    \label{fig:Bfield_ud_0}
\end{figure}

The moir\'e Hofstadter bands at \( u_d = 0 \) for Phases I and II are shown in Figs.~\ref{fig:Bfield_ud_0}(a) and Figs.~\ref{fig:Bfield_ud_0}(b).
These features extend the analysis of Ref.\,\cite{wang2024phase,kolar2024hofstadter} by incorporating second harmonic contributions and the effects of displacement fields. These spectra illustrate the fractal band structure that emerges near the \( K \) valley as a function of the normalized magnetic flux per moir\'e unit cell, \( \phi/\phi_0 \), for opposite orientations of the magnetic field applied perpendicular to the bilayer plane, where positive and negative flux correspond to magnetic fields along the $-z$  and $+z$ directions, respectively. We note that spectrum at the $K'$ valley is obtained by time-reversal operation, which flips the chirality of the electronic states and the direction of the magnetic field. A comment of the Zeeman effect is in order before discussing the spectrum. For $\theta=2.1^\circ$, the magnetic field strength needed to produce $\phi_0$ is approximately $25T$. The g-factor for bilayer MoTe2 varies from 4.73(at 2K) to 2.54(at 70K)\cite{jiang2017zeeman}. Thus, the Zeeman splitting at $\phi\approx \phi_0$ is estimated to be around $7$ meV, which is comparable to the first moir\'e band gap. Thus, with a  magnetic field of this magnitude, first moir\'e band will be valley-polarized, in which case it is legitimate to consider only one valley.

The evolution of the bands exhibits a characteristic dependence on the magnetic field orientation, particularly in the behavior of the gap between the two topmost moir\'e bands. 
Specifically, the gap decreases for positive flux and increases for negative flux.
{It's worth noting that, in our convention, a positive flux corresponds to a magnetic field pointing in the negative $\hat{\mathbf{z}}$-direction.}
This asymmetry correlates with the interplay between the external magnetic field direction and the average Berry curvature. In Phase I, the average Berry curvature of the top bands is positive, and the gap evolution closely mirrors that of conventional Landau levels. A similar trend is observed in Phase II, even though the top two bands exhibit opposite average Berry curvatures. 

While the spectra show in Figs.~\ref{fig:Bfield_ud_0} exhibit a wealth of intricate features, a particularly robust reorganization of states emerges near one flux quantum ($\phi/\phi_0 = 1)$ in both Phases I and II. Notably, small deviations from this flux value give rise to a well-defined sequence of Landau fans. As discussed in Sec. \ref{sec:CF states}, this structure signals the presence of a stable sequence of Jain FCI states when such Landau fans are associated with filled composite fermion bands. 

\subsection{Moir\'e Hofstadter bands at non-zero displacement field}
\label{sub:moir_e_hofstadter_bands_at_non_zero_displacement_field}
{

Beyond the moir\'e Hofstadter band structure, the electric response of the system to a perpendicular displacement field $u_d$ provides a tunable knob to probe and control quantum phase transitions. The displacement field $(u_d/2)\,\sigma_z$ in Eq.\eqref{eq: continuum model with B and E fields} acts on the layer degrees of freedom. In the large $u_d$ limit ($|u_d| \gg |T|$), the two layers become effectively decoupled, leading to Landau levels in each layer that are affected by periodic moir\'e potentials $V(\bs{r})$.

We first note the displacement field can 
alter the topological character of the moir\'e bands at $\phi=0$. As shown in 
Fig.~\ref{fig: displacement field 21} and Fig.~\ref{fig: displacement field 389}, increasing $u_d$ affects the bandwidths of the moir\'e bands leading to a sequence of phase transitions at $\phi=0$. In particular, the gap between the first and second bands closes at  $u_d \approx 17.0$ meV ($D/\varepsilon_0 \approx {u_d/(e\cdot d)\cdot\varepsilon_r}=242.9$ mV/nm) for $\theta \approx 2.1^{\circ}$, and at $u_d \approx 19.5$ meV ($D/\varepsilon_0 \approx 278.6$ mV/nm) for $\theta\approx3.89^{\circ}$. These estimates assume an inter-layer distance $d\approx 0.7$ nm{\cite{lin2014ambipolar,wang2019precise,zazpe20212d}} and a relative permittivity $\varepsilon_r\approx 10${\cite{laturia2018dielectric,kutrowska2022exploring}}. After the gap closing, the Chern number of the first moir\'e band becomes zero.

Moreover, the interplay between the displacement field and the external magnetic field gives rise to rich and intricate behavior. When \( u_d \) becomes comparable to the scales of the moir\'e potential and interlayer tunneling, nontrivial interlayer mixing broadens the moir\'e Hofstadter bands, as the displacement field shifts the two layers in opposite directions along the energy axis (see Fig.~\ref{fig: displacement field 21} and Fig.~\ref{fig: displacement field 389}). 
The spectrum displays a robust sequence of gaps as the displacement field increases, which occur in the top $p$ Hofstadter sub-bands at fluxes $\phi/\phi_0 = \frac{2p}{2p+1}$ and $\frac{2p+2}{2p+1}$, 
at both $\theta \approx 2.1^{\circ}$ and $3.89^{\circ}$.

In contrast, at several other fluxes, such as $\phi/\phi_0=8/5, 2/5, 4/9$, 
we identify topological quantum criticality induced by the displacement field. These quantum phase transitions at flux $\phi/\phi_0 = p/q$ are characterized by Chern number exchange $\Delta C = \pm q$ through $q$ Dirac cone band touchings, following the multiplicity enforced by the magnetic translation group~\cite{Lee2018Emergent,wang_classification_2020}. We further discuss this scenario in Sec.~\ref{sec:CF states} (showing the multiplicity of Dirac cones in Fig.~\ref{fig: QPT 21} and Fig.~\ref{fig: QPT 389}), highlighting the displacement field as an experimental control knob for topological quantum critical phenomena.

\begin{figure*}[!htb]
    \centering
    \includegraphics[width=1\linewidth]{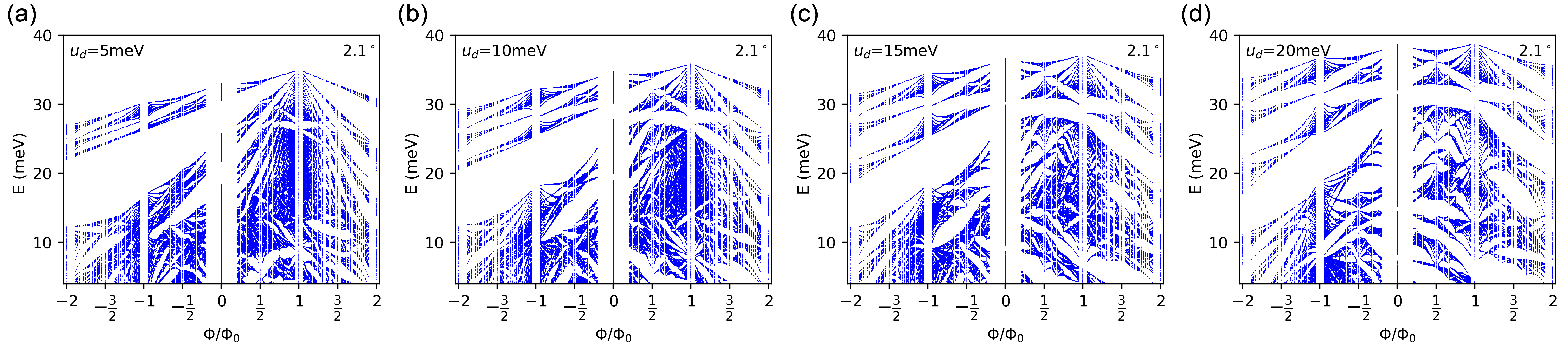}
    \caption{The moir\'e Hofstadter spectra at $2.1^{\circ}$ under increasing displacement field. (a-d)
    $u_d=5,10,15,20$ meV. The spectra broaden as inter-layer mixing is progressively suppressed.
    The first moir\'e band at $\Phi/\Phi_0=0$ becomes trivial at $u_d \approx 17$ meV [between (c) and (d))].}
    \label{fig: displacement field 21}
\end{figure*}

\begin{figure*}[!htb]
    \centering
    \includegraphics[width=1\linewidth]{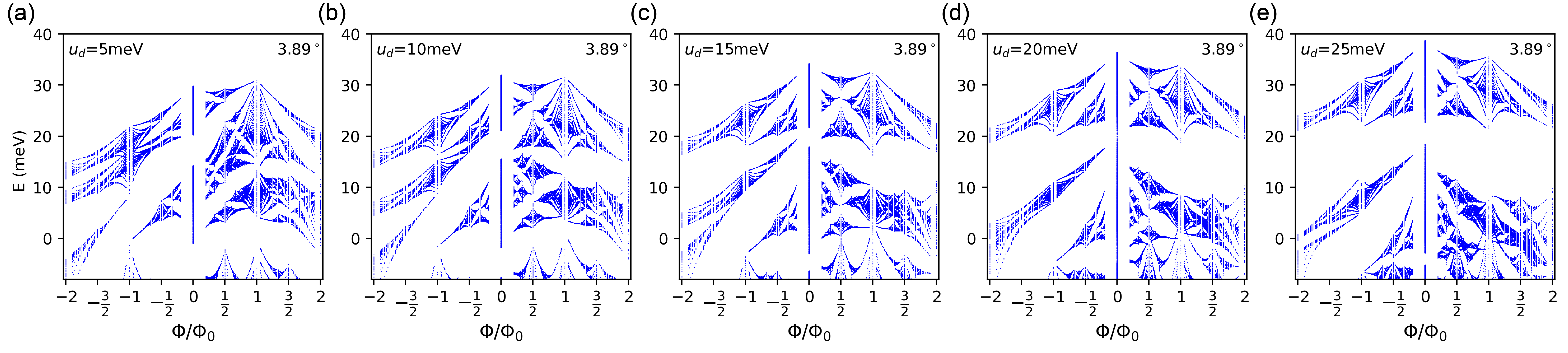}
    \caption{The moir\'e Hofstadter spectra at $3.89^{\circ}$ under increasing displacement field. (a-e)
    $u_d=5,10,15,25$ meV. The spectra broaden as inter-layer mixing is progressively suppressed. The first moir\'e band at $\Phi/\Phi_0=0$ becomes trivial at $u_d \approx 19.5$ meV [near (d)[.}
    \label{fig: displacement field 389}
\end{figure*}

\section{Composite fermion states: platforms for FCIs and FQSH} \label{sec:CF states}

The analysis of moir\'e Hofstadter spectrum discussed in Sec.~\ref{sec:bands with E and B fields} also provides a useful framework 
for understanding topological orders in t-MoTe$_2$, described via composite fermions through attachment of Chern-Simons statistical flux tubes onto particles.
Such a flux attachment is made possible via a singular gauge transformation, subject to a condition that the flux field $b$ is related to the local density through $b(\bm{r})=s\phi_0\rho(\bm{r})$, where $s$, the number of flux quanta attached to each particle, is even so that composite particles maintain fermionic character.

In traditional fractional quantum Hall systems, the flux attachment results in an effective integer quantum Hall state of composite fermions that fully filled ``Landau levels'' -- also known as Lambda levels~\cite{jain2007composite}.

The local density constraint can be enforced through a Chern-Simons (CS) gauge theory for the fluxes $b(\bm{r})$, with a uniform mean-field Chern-Simons field $\bar b=s\phi_0\bar\rho$ characterizing the quantum Hall liquid regime~\cite{lopez1991fractional}.

In the case of small angle twisted t-MoTe$_2$ that demonstrates multiple $C=\pm1$ moir\'e bands, the situation is topologically equivalent to Landau levels. There are two major differences though. One is that in t-MoTe$_2$ different valleys have opposite spin polarizations, and the other one is, in addition to the $k^2$-dispersion for the particles [c.f. Eq.\eqref{eq: polarized continuum model K valley}] the system also features a moir\'e periodicity. 

While the first point can be directly addressed by attaching opposite CS flux to opposite valley, the second one implies the tendency for a moir\'e periodically modulated CS field $b$.  
However, we argue that when the topological order is robust, a uniform \( b \) ansatz remains a valid starting point for characterizing the ground state's topological properties. In this regime, the primary effect of a non-uniform $b$ is to introduce dispersion into the composite fermion Lambda levels. Compared to a uniform field $\bar{b}$ of the same average strength, the spatially varying case potentially leads to a reduced gap, making the system more susceptible to perturbations. If interactions were to destabilize the Lambda levels altogether, the system would likely transition to a different correlated phase.

However, recent experimental observations of FCIs in t-MoTe\(_2\)
~\cite{cai2023signatures, park2023observation, zeng2023thermodynamic, xu2023observation, park2025ferromagnetism, park2025observation, xu2025signatures, redekop2024direct}
have revealed a hierarchy of incompressible states reminiscent of those in conventional FQH systems, despite the absence of an external magnetic field. These experiments not only report the observation of hierarchical Jain states at filling fractions \( \nu_h = 2/3, 3/5, 4/7, 5/9 \), but also find gaps that are substantially larger~\cite{redekop2024direct} than those typically observed in traditional FQH systems.

Informed by this phenomenology, we henceforth investigate a uniform Chern-Simons (CS) flux configuration, which we expect to be adiabatically connected to a self-consistent, spatially non-uniform CS field. In this regime, the uniform CS field serves as a good approximation for describing the most robust topologically ordered states. In contrast, the more fragile Lambda levels obtained from the uniform \( b \) ansatz may require a more refined treatment that self-consistently accounts for spatial variations in the CS field, an important direction for future work that lies beyond the scope of this study. Below, we present explicit examples illustrating both robust and fragile composite fermion gaps under a uniform \( b \) approximation, and characterize their topological properties in the context of both valley-spin-polarized FCIs and time-reversal-invariant FQSH phases described by valley-contrasting CS flux attachment.

\begin{figure*}[!htb]
    \centering
    \includegraphics[width=1\linewidth]{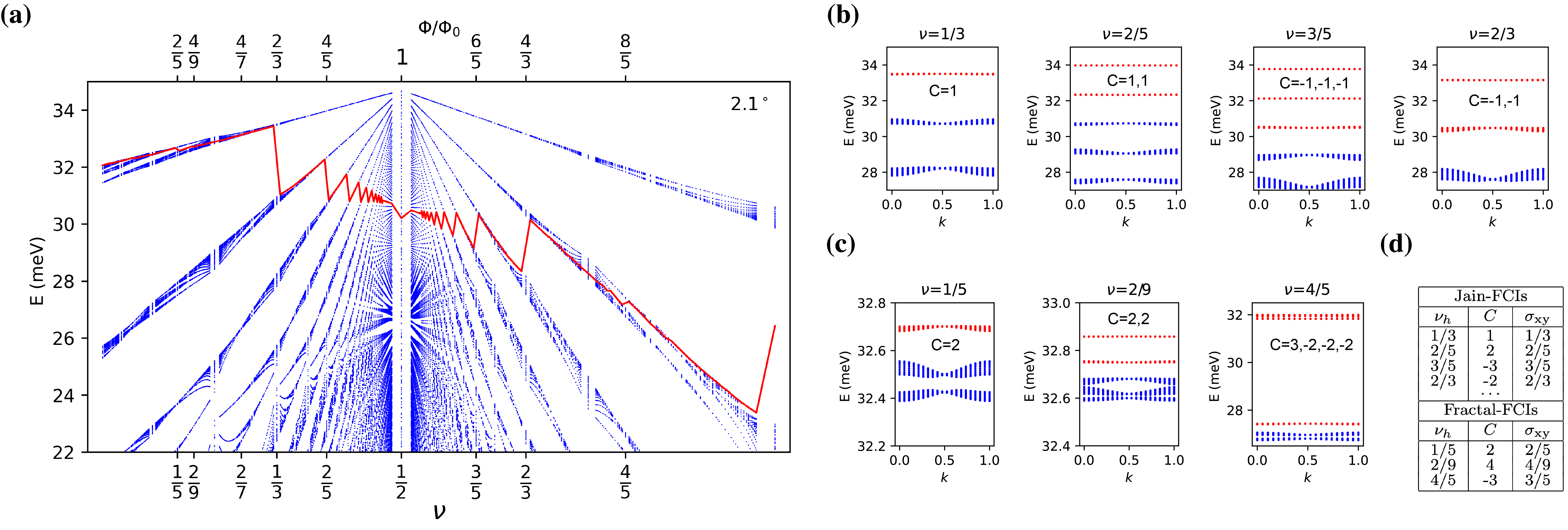}
    \caption{(a) Hofstadter spectrum at $2.1^{\circ}$ and $u_d=0$. The fermi surface (red curve) exhibits incompressible CF states at various filling fractions, indicated by abrupt jumps. (b) The moir\'e Hofstadter bands of Jain-FCIs. (c) The moir\'e Hofstadter bands of Fractal-FCIs. (d) Table summarizing the Hall conductance of Jain-FCIs and Fractal-FCIs.}
    \label{fig: Jain/Fractal 21}
\end{figure*}

\begin{figure*}[!htb]
    \centering
    \includegraphics[width=1\linewidth]{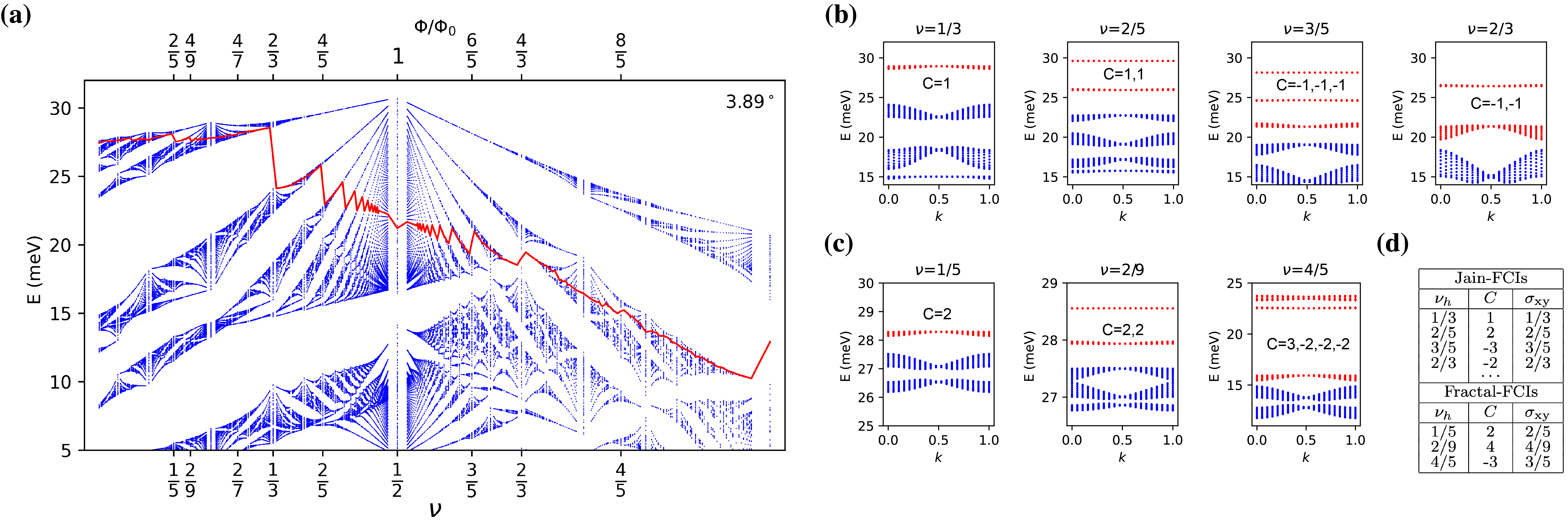}
    \caption{(a) Hofstadter spectrum at $\theta\approx3.89^{\circ}$ and $u_d=0$. The fermi surface (red curve) exhibits incompressible CF states at various filling fractions, indicated by abrupt jumps. (b) The moir\'e Hofstadter bands of Jain-FCIs. (c) The moir\'e Hofstadter bands of Fractal-FCIs. (d) Table summarizing the Hall conductance of Jain-FCIs and Fractal-FCIs.}
    \label{fig: Jain/Fractal 389}
\end{figure*}

We investigate composite fermions characterized by two CS flux quanta per 
composite fermion ($s=2$) \cite{jain1989composite,lopez1991fractional,Kol1993,Moller2015,murthyshankar2012,Sohal-2018,wang_classification_2020,lu_santos_2024fractional}. While this is known to give rise to the Jain sequence at filling 
$\nu_{\text{Jain}}=\frac{1}{3},\frac{2}{5},...$ in Landau level systems \cite{jain1989composite,lopez1991fractional}, we investigate the onset of composite fermion states in the context of the continuum model of t-MoTe$_2$.

As discussed above, we attach uniform $b$ to $K$-valley Hamiltonian Eq.~\eqref{eq: continuum model with B and E fields}, and $-b$ to $K'$-valley Hamiltonian. Given a particular field parametrized as $\phi=b A_{uc}=\frac{p}{q}\phi_0$ with $p$ and $q$ coprime integers (same as in Sec.\ref{sec:bands with E and B fields}), the moir\'e hole filling fraction per valley is given by 
\begin{equation}
    \nu=\frac{1}{2}\frac{\phi}{\phi_0}=\frac{1}{2}\frac{p}{q}.\label{eq:fillingpq}
\end{equation}
This indicates that for each chosen value of $\phi$, one can trace the Fermi energy by checking the filling factor using Eq.\eqref{eq:fillingpq}, and by doing so the composite fermion gap at 
{$\nu$} 
and other fillings can be directly extracted. 
Moreover, the filling factor of the corresponding composite fermions reads
\begin{equation}
\label{eq: CF filling main}
    \nu_\text{CF}=\frac{p}{2},
\end{equation}
see Appendix \ref{sec:appCSFluxAttachment} for more details. For $p\in$ even integer, there are $\nu_\text{CF}$ fully filled composite fermion bands, for which we can calculate the corresponding Chern numbers {$C$}, which are related to the
fractional Hall conductivity as {\cite{Moller2015}}

\begin{equation}
\label{eq: sigma xy}
{
\sigma_{xy} = \frac{e^2}{h} \frac{C}{2C + 1}
}
\,.
\end{equation}
The methodology underlying Eqs.~\eqref{eq:fillingpq}, \eqref{eq: CF filling main}, and \eqref{eq: sigma xy} provides a framework for characterizing the emergent topological orders within the composite fermion theory.

In Fig.\ref{fig: Jain/Fractal 21} we present the composite fermion bands for $\theta=2.1^\circ$ t-MoTe$_2$, using the uniform flux approximation outlined above. Under this assumption, the Hofstadter spectrum in Fig.\ref{fig: Jain/Fractal 21}(a) is obtained in  the same way as in Sec. \ref{sec:bands with E and B fields}, with the red curve being the Fermi energy indicator. By tracking
the jumps of the Fermi level, we find two distinct gapped states: the usual Jain states at $\nu=\nu_\text{Jain}$, and a series of fractal FCI states at a different set of fillings, which are shown in Fig.\ref{fig: Jain/Fractal 21}(b) and (c) respectively. Note that for the Jain states, the filled Lambda levels (in red) are almost flat and well-separated. Each of the band carries Chern number $C=\pm1$ depending on whether $\nu<\frac{1}{2}$ or $\nu>\frac{1}{2}$. These features imply robust FCI states and validate the uniform flux approximation {\it a posteriori}.
For the other fractal FCI states, the gaps are smaller and each filled composite band carries a higher Chern number $C=2$ or $C=3$. 
{Examples of such states are shown in Fig. \ref{fig: Jain/Fractal 21} at filling fractions $\nu = 4/5, 2/9, 1/5$, characterized by Hall conductance $\sigma_{xy} = (3/5)(e^2/h), (4/9)(e^2/h), (2/5)(e^2/h)$, respectively.}
Compared to the Jain states, these states are more fragile and may not be stable as $\nu$ gets close to $\frac{1}{2}$ and the gap becomes even smaller. A summarizing list of these two FCI states is shown in Fig.\ref{fig: Jain/Fractal 21}(d).

Similar analysis of composite fermion bands for $\theta=3.89^\circ$ t-MoTe$_2$ is shown in Fig.~\ref{fig: Jain/Fractal 389}. Here we have also identified both the Jain and fractal FCI states, and the features are similar to those in Fig.~\ref{fig: Jain/Fractal 21}. The noticeable differences imply that as $\theta$ increases, the composite fermion bands become slightly more dispersive. Despite this difference, we note a common feature in both phases that as $\nu$ approaches $\frac{1}{2}$, the system becomes a gapless composite Fermi liquid \cite{PhysRevLett.131.136501,PhysRevLett.131.136502}.

Extending the analysis to the second moir\'e bands is straightforward. As we show in the Appendix \ref{sec:secondband}, the CF gap for the second moir\'e bands are quite small compared to those in the first moir\'e bands, indicating more fragile topological orders in the second bands. This is the consistent with experiments, {showing} that the most salient FCI states observed so far are all from partially filling the first moir\'e band~\cite{cai2023signatures, park2023observation, zeng2023thermodynamic, xu2023observation, park2025ferromagnetism, park2025observation, xu2025signatures, redekop2024direct}.

So far what we have discussed are the spin or valley polarized FCI states. In principle, by attaching opposite fluxes to different valleys, we are able to describe topological orders with time-reversal invariance, i.e. the fractional quantum spin Hall states or fractional topological insulators, also in the composite fermion picture. The above results are directly applicable, meaning that both the Jain and fractal FTI states can may exist in this system. However, unlike the FCI states where the redidual interaction in the composite fermion picture can be neglected, in the CF picture of FTI states the inter-valley interaction is not negligible and usually leads to time-reversal symmetry breaking orders~\cite{kang2025time,park2025ferromagnetism}.

The composite fermion picture also allows for a  new analysis on how the external $E$ field affects the FCI phases in t-MoTe$_2$. To this end, we follow the recipe outlined in Sec.\ref{sub:moir_e_hofstadter_bands_at_non_zero_displacement_field} and investigate the composite fermion spectrum in the presence of $E$ field. 
{We restrict to values of the field for which the first moir\'e band remains topological, which corresponds to $0 < u_d < 17$ meV for $\theta = 2.1^{\circ}$ and $0 < u_d < 19.5$ meV for $\theta = 3.89^{\circ}$.}
The results for $\theta=2.1^\circ$ and $\theta=3.89^\circ$ are presented in Fig.~\ref{fig: QPT 21} and Fig.~\ref{fig: QPT 389} respectively. In both cases, we find a common feature that the FCI gaps tend to decrease as the electric field increases. 

\begin{figure}[!htb]
    \centering
    \includegraphics[width=1\linewidth]{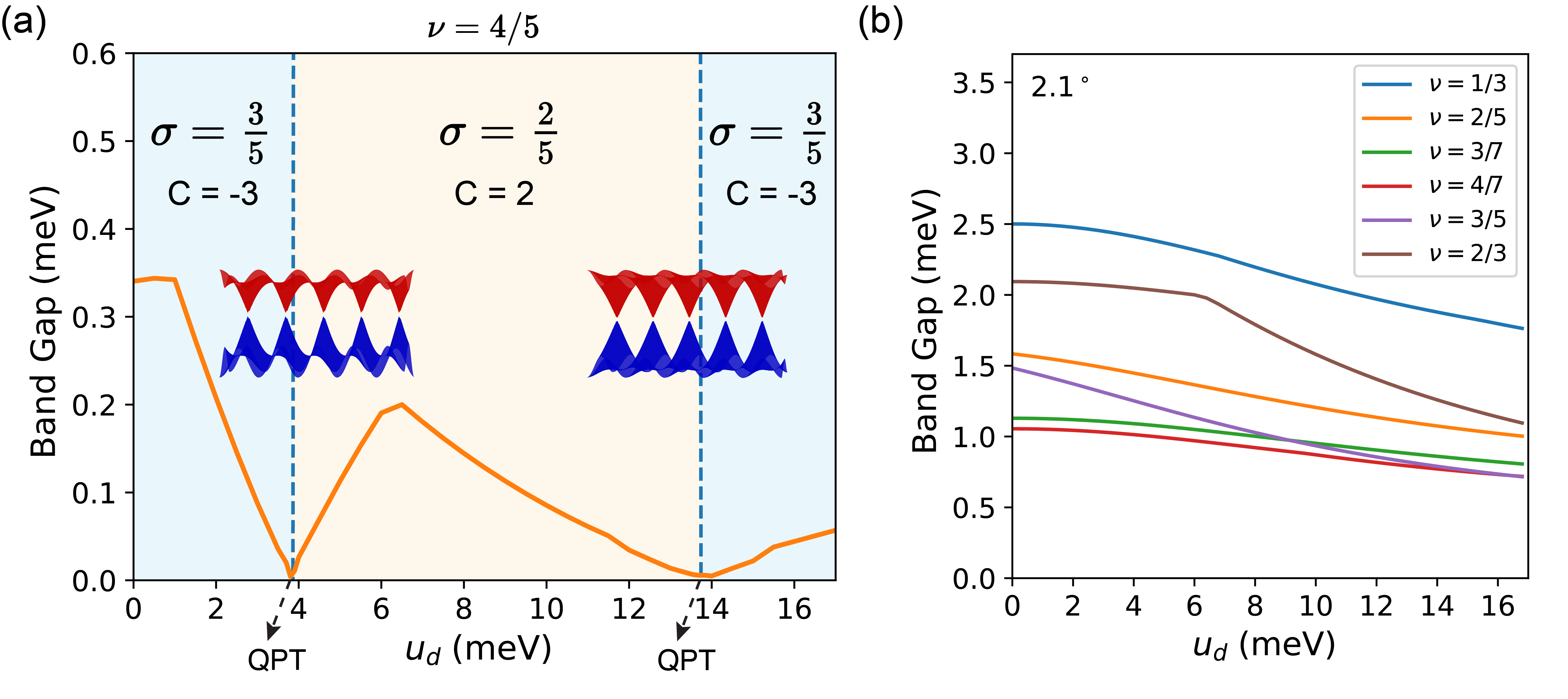}
    \caption{Composite fermion band gap at $2.1^{\circ}$. (a) At filling $\nu=\frac{4}{5}$ (Fractal-FCI), two quantum phase transitions occur, during which the occupied band (red) and the unoccupied band (blue) exchange Chern number through five Dirac cones. (b) For Jain-FCIs, the composite fermion band gaps remain open, demonstrating their robustness. }
    \label{fig: QPT 21}
\end{figure}

\begin{figure}[!htb]
    \centering
    \includegraphics[width=1\linewidth]{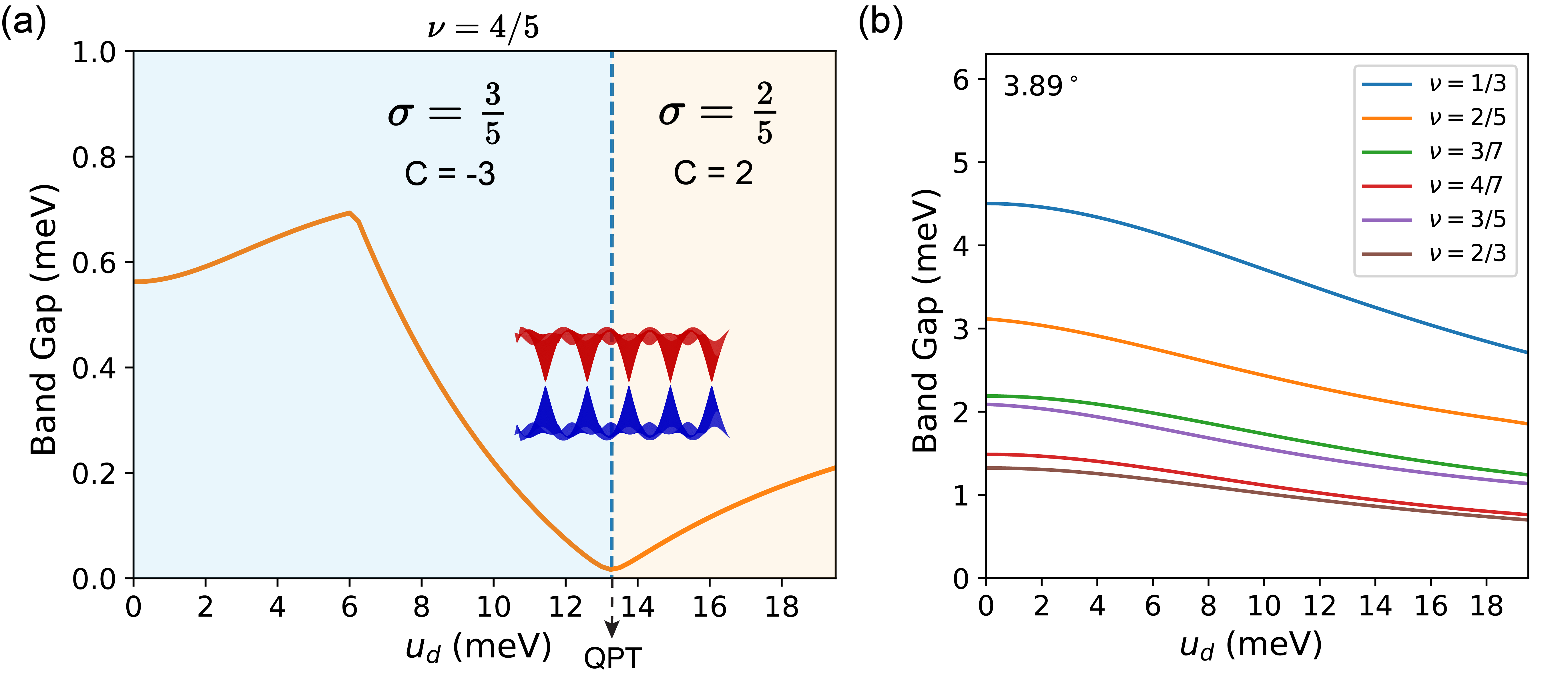}
    \caption{Composite fermion band gap at $3.89^{\circ}$. (a) At filling $\nu=\frac{4}{5}$ (Fractal-FCI), one quantum phase transitions occurs, where the occupied band (red) and the unoccupied band (blue) exchange Chern number through five Dirac cones. (b) For Jain-FCIs, the composite fermion band gaps remain open, demonstrating their robustness. }
    \label{fig: QPT 389}
\end{figure}

This feature underscores the role of interlayer tunneling in stabilizing FCI states, which is suppressed by the applied electric field. Such suppression of composite fermion gaps could be observed experimentally through variations in the thermal activation gaps across FCI plateaus as a function of the electric field.

While the Jain states remain gapped with $u_d$ up to $17$meV, {remarkably,} we identified an electric field induced quantum phase transition (QPT) in the fractal FCI states. At $\nu=\frac{4}{5}$ and $\theta=2.1^\circ$, such QPT occurs twice as $u_d$ increases from $0$ to around $17$meV, as demonstrated in Fig.\ref{fig: QPT 21}(a).
{The first transition between FCIs with $\sigma_{xy} = (3/5)(e^2/h)$ and $\sigma_{xy} = (2/5)(e^2/h)$
occurs at $u_d \approx 4$ meV, and a second transition back to $\sigma_{xy} = (3/5)(e^2/h)$ takes place at $u_d \approx 14$ meV. 
}
For $\nu=\frac{4}{5}$ and $\theta=3.89^\circ$, such a QPT only occurs once when $u_d$ sweeps from $0$ to $19.5$meV, as shown in Fig.\ref{fig: QPT 389}(a). 

As shown in the figure insets, these quantum phase transitions are marked by a Chern number transfer of $\Delta C = \pm 5$ between filled composite fermion bands, indicating the emergence of five composite Dirac fermions. This result demonstrates that exotic multi-flavor quantum critical points \cite{Lee2018Emergent,wang_classification_2020} can be accessed by tuning the electric field in dual-gated moir\'e systems. In the Appendix \ref{sec:moreQPT}, we provide supplemental plots for QPTs at fractal FCIs at filling fraction $2/9$ and $1/5$.

\section{Discussion and Outlook}
\label{sec:discussion}

In this work we have investigated the fractal electronic spectrum of t-MoTe$_2$ in the presence of both out-of-plane magnetic field and electric field, for two particular phases with different twist angles: $\theta\approx 2.1^\circ$ and $\theta\approx 3.89^\circ$.  Our approach starts with the continuum model description and includes both the first and second harmonics for the intralayer moir\'e potential and interlayer tunneling. Given a particular valley, the moir\'e band gap displays distinct behavior for positive flux and negative flux, reflecting the nonzero Chern number character of each moir\'e band. In addition, we also observe a Landau fan structure stemming from $\phi=\phi_0$ in both phases.
Introducing electric field tends to broaden each moir\'e band even in the absence of magnetic field such that, at large enough electric field, the gap between the first two moir\'e bands closes and the two bands become trivial, {a regime that appears to be within experimental reach.} 

For $\theta\approx 2.1^\circ$, we obtain for the displacement field a critical value of $17.0$ meV, while for $\theta\approx 3.89^\circ$ we obtain a larger critical value of $19.5$ meV. 

We have also shown that the analysis of 
the fractal spectrum in the presence of magnetic field is of particular significance in {characterizing}
the composite fermion 
{states describing a host of}
FCI states in t-MoTe$_2$. 
{In particular, we have 
done a comprehensive analysis of the
Hofstadter spectrum of composite fermions
associated with a uniform Chern-Simons field that attaches two flux quanta per particle.}
We have identified
both Jain FCI states at filling fractions $\nu=\nu_\text{Jain}=\frac{1}{3},\frac{2}{5},...,\frac{3}{5},\frac{2}{3}$, and a series of fractal FCI states at filling fractions $\nu=\nu_\text{Fractal}=\frac{1}{5},\frac{2}{9},\frac{4}{5},...$. Although the composite fermion bands for the Jain states are similar to the Landau levels with Chern numbers $C=\pm1$, those for the fractal FCI states are more dispersive, with Chern numbers $C=2$ or $3$, which do not resemble Landau levels. 
{Our analysis reveals}
that Jain states generally exhibit larger composite fermion gaps, 
{indicating greater stability.}
{
While we have primarily focused on filling fractions $\nu<1$, where prominent composite fermion gaps are observed, an example for $1<\nu<2$ at $\theta\approx 2.1^{\circ}$ is discussed in Appendix~\ref{sec:secondband}.}

{Moreover, we show that an perpendicular electric field suppresses the composite fermion gaps, and 
in some cases, can tune the FCI through topological quantum phase transition.}
Investigating the fractal FCI states and their quantum phase transitions in future experiments will certainly enrich our understanding of topological orders in moir\'e materials.

{Besides the displacement field, temperature plays a critical role in observing the Hofstadter spectrum and the stability of the associated correlated phases. 
In moir\'e systems under external magnetic fields, both single-particle and fractionalized states associated with the fractal spectrum have been experimentally observed at temperatures ranging from 10 mK to 4.5 K\cite{Dean13, Ponomarenko13, Hunt13, Forsythe18, Spanton18, das2021observation,das2021symmetry,BalentsEfetovYoung20,stepanov2021competing,Saito21,wu2021chern,xie2021fractional}.  On the other hand, in the absence of external field, recent experiments\cite{park2023observation,redekop2024direct,park2025ferromagnetism} on FCIs in t-MoTe$_2$ report thermal activation gaps of $\Delta_{\nu=2/3}=23$ K and $\Delta_{\nu=3/5}=15$ K at $\theta=3.7^{\circ}$\cite{park2023observation}, and $\Delta_{\nu=2/3}=$ 27 K at $\theta=3.64^{\circ}$\cite{redekop2024direct}.
For comparison, our composite fermion analysis at $\theta=3.89^{\circ}$ yields band gaps of 1.3 meV for $\nu=2/3$ and 2.1 meV for $\nu=3/5$, corresponding to thermal energy gaps of approximately $\Delta_{\nu=2/3}\approx 15$ K and $\Delta_{\nu=3/5}\approx 24$ K respectively, which are consistent with the experimentally reported scales.}

Our discussion of Hofstadter spectrum of t-MoTe$_2$ also serves as a departing point for future studies on the interplay between electron correlation and external magnetic field in this platform. 
{A compelling question is to investigate how interacting symmetry breaking electronic orders, nested in the 
Hofstadter system, evolve and compete under large magnetic fluxes.}
Another open question involves the treatment of the Chern-Simons statistical flux field $b(\bm{r})$ in a self-consistent way. We have argued that for a robust FCI state, approximating $b(\bm{r})$ as some uniform $\bar b$ does not change the essential physics. In this situation, electron interaction stabilizes the FCI state and the composite fermions are treated as non-interacting particles. %
{However, this uniform field ansatz is biased towards liquid-like FCI states and thus may underestimate the role played by other electronic orders competing with the FCI state\cite{tevsanovic1989hall,Kol1993}.
{This is particularly relevant for the $\nu=1/3$ state,
which emerges as one of the prominent gaps in our composite fermion analysis, but has not yet been observed experimentally. 
Several experimental studies on moir\'e systems, including t-MoTe$_2$, suggest that the observed 1/3 state is likely a trivial correlated insulator\cite{xu2025signatures,xu2020correlated,huang2021correlated}. 
This discrepancy may reflect the assumption inherent in our uniform flux attachment approximation, which favors fractional states over competing topologically trivial orders. A self-consistent treatment of the Chern-Simons flux attachment, allowing for spatial variations in density, could provide further insight into the competition between FCIs and CDWs in this regime.} Understanding these non-uniform flux-charge binding states is an outstanding question that warrants future investigation.
}

\section*{Acknowledgments}
We acknowledge useful discussions with Andrei Bernevig, Claudio Chamon, Sankar Das Sarma, Jainendra Jain, Steven Kivelson, Weijie Li, Fengcheng Wu, Sri Raghu, Nicolas Regnault, and Ajit Srivastava. 
This research was supported by the U.S. Department of Energy, Office of Science, Basic Energy Sciences, under the DE-SC0023327 award (T. L. and L. H. S.), and by the Gordon and Betty Moore Foundation's EPiQS Initiative through GBMF8686 (Y.-M. W).

\appendix
\section{Hamiltonian in the Landau Level basis}

Following \cite{murthy2003hamiltonian}, we perform flux attachment to electrons characterized by $\boldsymbol{\Pi} = \mathbf{p}+e \mathbf{A}$ in an external magnetic field $\mathbf{B}=\nabla \times \mathbf{A}=- B \hat{\mathbf{z}}$.

The cyclotron coordinate is defined as 
\begin{equation}
    \boldsymbol{\eta}=\ell^2 \hat{\mathbf{z}} \times \boldsymbol{\Pi}\quad \Leftrightarrow\quad \begin{cases}
        \eta_x = -\ell^2\Pi_y\\
        \eta_y = \ell^2 \Pi_x
    \end{cases},
\end{equation}
where $\ell^2=\hbar / e B$.

The guiding center coordinate is defined as
\begin{equation}
    \mathbf{R}=\mathbf{r}-\boldsymbol{\eta}.
\end{equation}

The cyclotron and guiding center coordinates follow such commutation relations
\begin{equation}
    \left[\eta_x, \eta_y\right]=i \ell^2,\quad
    \left[R_x, R_y\right]=-i \ell^2,
\end{equation}
and the ladder operators for $\boldsymbol{\eta}$ can be defined as
\begin{equation}\label{LLladder}
\begin{split}
&a=\frac{\eta_x+i \eta_y}{\sqrt{2 \ell^{ 2}}}, \quad a^{\dagger}=\frac{\eta_x-i \eta_y}{\sqrt{2 \ell^2}},
\end{split}
\end{equation}
where $\left[a, a^{\dagger}\right]=1$ and $\hat{a},\hat{a}^{\dagger}$ define a set of Landau levels $|n\rangle$.

We aim to express the operator $V_{\bs{g}} \equiv e^{i \bs{g} \cdot \bs{r}}$ in the Landau level basis. We begin with writing $\bs{r} = \bs{R}+\bs{\eta}$, which allows us to decompose $V_{\bs{g}}  =  e^{i \bs{g} \cdot \bs{\eta}}\,e^{i \bs{g} \cdot \bs{R}}=\mathcal{A}(\boldsymbol{g})\mathcal{B}(\boldsymbol{g})$ using the fact that the cyclotron and guiding center coordinates commute. Since each $\boldsymbol{g}$ in Eq.\eqref{1stHarmonics} and Eq.\eqref{2ndHarmonics} 
is a linear combination of the primitive vectors, we choose to expand them in terms of $\boldsymbol{g}_1$ and $\boldsymbol{g}_2$ as $\boldsymbol{g} = m\boldsymbol{g}_1+n\boldsymbol{g}_2$, and thus express $V_{\boldsymbol{g}}$ as
\begin{equation}
    V_{\boldsymbol{g}} = e^{i (m\boldsymbol{g}_1+n\boldsymbol{g}_2) \cdot \bs{\eta}}\,e^{i (m\boldsymbol{g}_1+n\boldsymbol{g}_2) \cdot \bs{R}} = \mathcal{A}_{m,n}\mathcal{B}_{m,n}.
\end{equation}

Here we summarize $m,n$ for every $\boldsymbol{g}$ vector that appears in the moir\'e potentials and tunneling terms, in Table.\,\ref{mntable}.

\begin{table}[!htb]
\centering
\begin{tabular}{c @{\hspace{0.3cm}} r r @{\hspace{0.8cm}} c @{\hspace{0.4cm}} r r}
\toprule
\textbf{Index} & \textbf{$m$} & \textbf{$n$} & \textbf{Index} & \textbf{$m$} & \textbf{$n$} \\
\midrule
$\vec{g_1}$ & 1  & 0   & $\vec{g_6}$           & 1   & -1 \\
$\vec{g_2}$ & 0  & 1   & $\vec{g_1}+\vec{g_2}$ & 1   & 1  \\
$\vec{g_3}$ & -1 & 1   & $\vec{g_3}+\vec{g_4}$ & -2  & 1  \\
$\vec{g_4}$ & -1 & 0   & $\vec{g_5}+\vec{g_6}$ & 1   & -2 \\
$\vec{g_5}$ & 0  & -1  & $\vec{g_2}+\vec{g_3}$ & -1  & 2  \\
\bottomrule
\end{tabular}
\caption{Integer coefficient pairs $(m, n)$ for the vectors $\boldsymbol{g}_i$ and their combinations.}
\label{mntable}
\end{table}

For a generic $\boldsymbol{q}$, the matrix element in the Landau level basis $\left\langle n_2\right| \mathcal{A}(\pm\boldsymbol{q})\left|n_1\right\rangle=\left\langle n_2\right| e^{\pm i \mathbf{q} \cdot \boldsymbol{\eta}}\left|n_1\right\rangle$ is 
\begin{equation}
\label{SM: Laguerre}
\begin{split}
    \left\langle n_2\right| e^{-i \mathbf{q} \cdot \boldsymbol{\eta}}\left|n_1\right\rangle&=\sqrt{\frac{n_{2}!}{n_{1}!}} e^{-x / 2}\left(\frac{-i q_{-} \ell}{\sqrt{2}}\right)^{n_1-n_2} L_{n_2}^{n_1-n_2}(x),\quad\\ 
    \left\langle n_2\right| e^{+i \mathbf{q} \cdot \boldsymbol{\eta}}\left|n_1\right\rangle&=\sqrt{\frac{n_{2}!}{n_{1}!}} e^{-x / 2}\left(\frac{i q_{-} \ell}{\sqrt{2}}\right)^{n_1-n_2} L_{n_2}^{n_1-n_2}(x),
\end{split}
\end{equation}
where $q_- = q_x -iq_y,\, x = |q_-|^2\ell^2/2$, $n_1 \ge n_2$ and $L_{n_2}^{n_1-n_2}(x)$ is the Laguerre polynomial
\begin{equation}
    L_{n_2}^{n_1-n_2}(x)=\sum_{t=0}^{n_2} \frac{n_{1}!}{\left(n_2-t\right)!\left(n_1-n_2+t\right)!} \frac{(-1)^t}{t!} x^t\,.
\end{equation}

Instead of using guiding center coordinates along the $x$- and $y$-axes, i.e., $R_x = \vec{R}\cdot\hat{x},\,R_y = \vec{R}\cdot\hat{y}$, we project the guiding center $\boldsymbol{R}$ onto the reciprocal vectors $\boldsymbol{g}_1$ and $\boldsymbol{g}_2$ by defining
\begin{equation}
    R_1 = \hat{\boldsymbol{g}}_1\cdot\boldsymbol{R}\,,\quad R_2 = \hat{\boldsymbol{g}}_2\cdot\boldsymbol{R}\,,
\end{equation}
which satisfies
\begin{equation}
\begin{split}
&[R_1, R_2] = -i\ell^{2}_{B} (\hat{\bs{g}}_1 \times \hat{\bs{g}}_2)\cdot\hat{z} = -i\frac{\sqrt{3}}{2}\ell_B^2\,,\\
&e^{-i g\,R_1}\,R_2 \,e^{i g\,R_1}
=R_2 + (-i g)[R_1, R_2] \\&\hspace{2.4cm}= R_2 - \frac{\sqrt{3}}{2}g\,\ell^{2}_{B}\,.
\end{split}
\end{equation}

Define $|n,y\rangle$ to be the eigenstate of the $R_2$ operator, we have
\begin{equation}
\begin{split}
e^{igR_2}|n,y\rangle &= e^{igy}|n,y\rangle,\\
e^{igR_1}|n,y\rangle &= |n,y-\Delta\rangle,
\end{split}
\end{equation}
where $\Delta = \frac{\sqrt{3}}{2}g\ell_B^2 \text{ and } g\Delta = \frac{2\pi q}{p}$, which defines a p-periodicity.

Using the Baker–Campbell–Hausdorff formula, we obtain
\begin{equation}
\begin{split}
\mathcal{B}_{m,n} &= e^{i(m\boldsymbol{g}_1+n\boldsymbol{g}_2)\cdot\boldsymbol{R}} = e^{imgR_1+ingR_2} \\
&= e^{-\frac{1}{2}[imgR_1,ingR_2]}e^{imgR_1}e^{ingR_2}
    \\&=e^{\frac{1}{2}mng^2[R_1,R_2]}e^{imgR_1}e^{ingR_2}\\
    &=e^{-i\frac{1}{2}mng\Delta}e^{imgR_1}e^{ingR_2}
    \\&=e^{-imn\frac{\pi q }{p}}e^{imgR_1}e^{ingR_2}\,.
\end{split}
\end{equation}

Writing $y = y_0 + (up+j)\Delta$, we have
\begin{equation}\label{SM: Bmn}
\begin{split}
&\mathcal{B}_{m,n}|n,y_0+(up+j)\Delta\rangle \\
=& 
    e^{-imn\frac{\pi q }{p}}e^{ing(y_0+(up+j)\Delta)}|n,y_0+(up+j-m)\Delta\rangle\\
    =&e^{ingy_0}e^{i\frac{2\pi q}{p}n(j-\frac{m}{2})}|n,y_0+(up+j-m)\Delta\rangle.
\end{split}
\end{equation}

Applying Eq.\eqref{SM: Bmn} to $\boldsymbol{g}$ in Table.\,\ref{mntable}, for the 1st harmonics in Eq.\eqref{1stHarmonics}, leads to
\begin{equation}
\begin{split}
\mathcal{B}_{1,0}|n,y\rangle&=e^{i\boldsymbol{g}_1\boldsymbol{R}}|n,y\rangle  = e^{igR_1}|n,y\rangle \\
&= |n,y_0+(up+j-1)\Delta\rangle\,,\\
\mathcal{B}_{0,1}|n,y\rangle&=e^{i\boldsymbol{g}_2\boldsymbol{R}}|n,y\rangle = e^{igR_2}|n,y\rangle \\&= e^{igy_0}e^{i2\pi\frac{q}{p}j}|n,y_0+(up+j)\Delta\rangle\,,
\end{split}
\end{equation}

\begin{equation}
\begin{split}
\mathcal{B}_{-1,1}|n,y\rangle&=e^{i\boldsymbol{g}_3\boldsymbol{R}}|n,y\rangle = e^{-igR_1+igR_2}|n,y\rangle \\&=e^{igy_0}e^{i2\pi\frac{q}{p}(j+1/2)}|n,y_0+(up+j+1)\Delta\rangle\,,\\
\mathcal{B}_{0,-1}|n,y\rangle&=e^{i\boldsymbol{g}_5\boldsymbol{R}}|n,y\rangle = e^{-igR_2}|n,y\rangle \\&=e^{-igy_0}e^{-i2\pi\frac{q}{p}j}|n,y_0+(up+j)\Delta\rangle\,,
\end{split}
\end{equation}

and for 2nd harmonics in Eq.\eqref{2ndHarmonics}, leads to
\begin{equation}
\begin{split}
\mathcal{B}_{1,1}|n,y\rangle&=e^{i(\boldsymbol{g}_1+\boldsymbol{g}_2)\boldsymbol{R}}|n,y\rangle  = e^{igR_1+igR_2}|n,y\rangle \\&
= e^{igy_0}e^{i\frac{2\pi q}{p}(j-\frac{1}{2})}|n,y_0+(up+j-1)\Delta\rangle\,,\\
\mathcal{B}_{-2,1}|n,y\rangle&=e^{i(\boldsymbol{g}_3+\boldsymbol{g}_4)\boldsymbol{R}}|n,y\rangle = e^{-i2gR_1+igR_2}|n,y\rangle \\&= e^{igy_0}e^{i\frac{2\pi q}{p}(j+1)}|n,y_0+(up+j+2)\Delta\rangle\,,\\
\mathcal{B}_{1,-2}|n,y\rangle&=e^{i(\boldsymbol{g}_5+\boldsymbol{g}_6)\boldsymbol{R}}|n,y\rangle = e^{igR_1-i2gR_2}|n,y\rangle \\&=e^{-i2gy_0}e^{i\frac{2\pi q}{p}(-2)(j-\frac{1}{2})}|n,y_0+(up+j-1)\Delta\rangle\,,\\
\mathcal{B}_{-1,2}|n,y\rangle&=e^{i(\boldsymbol{g}_2+\boldsymbol{g}_3)\boldsymbol{R}}|n,y\rangle = e^{-igR_1+i2gR_2}|n,y\rangle \\&=e^{i2gy_0}e^{i\frac{2\pi q}{p}2(j+\frac{1}{2})}|n,y_0+(up+j+1)\Delta\rangle\,,\\
\mathcal{B}_{-1,0}|n,y\rangle&=e^{i\boldsymbol{g}_4\boldsymbol{R}}|n,y\rangle = e^{-igR_1}|n,y\rangle \\&=|n,y_0+(up+j+1)\Delta\rangle\,.
\end{split}
\end{equation}

Now, combine $\mathcal{A}_{m,n}$ and $\mathcal{B}_{m,n}$, for any $m\boldsymbol{g}_1+n\boldsymbol{g}_2$, we have
\begin{equation}\label{Vmn}
\begin{split}
&V_{m\boldsymbol{g}_1+n\boldsymbol{g}_2}|n_1,y\rangle \\
&= e^{i(m\boldsymbol{g}_1+n\boldsymbol{g}_2)\cdot\boldsymbol{r}}|n_1,y\rangle =  \mathcal{A}_{m,n}\mathcal{B}_{m,n}|n_1,y\rangle\\
    &=e^{ingy_0}e^{i\frac{2\pi q}{p}n(j-\frac{m}{2})}\mathcal{A}_{m,n}|n_1,y-m\Delta\rangle\\
    &=e^{ingy_0}e^{i\frac{2\pi q}{p}n(j-\frac{m}{2})}\sum_{n_2}
    \langle n_2 ,y-m\Delta|
    \mathcal{A}_{m,n}|n_1,y-m\Delta\rangle\\&\hspace{5cm}\times|n_2,y-m\Delta\rangle\\
    &=e^{ingy_0}e^{i\frac{2\pi q}{p}n(j-\frac{m}{2})}\sum_{n_2}
    F_{n_2,n_1}(m\boldsymbol{g}_1+n\boldsymbol{g}_2)|n_2,y-m\Delta\rangle\,,
\end{split}
\end{equation}
where $F_{n_2,n_1}(m\boldsymbol{g}_1+n\boldsymbol{g}_2) = \langle n_2|e^{i(m\boldsymbol{g}_1+n\boldsymbol{g}_2)\cdot\boldsymbol{\eta}}|n_1\rangle$ is defined in Eq.\eqref{SM: Laguerre}

Besides, the parabolic terms are:
\begin{equation}
\begin{split}
    &-\frac{\hbar^2}{2m^*}(\boldsymbol{\Pi}-\boldsymbol{k}_{t/b})^2 \\=& 
    -\frac{\hbar^2}{2m^*}\left(
    \Pi_x^2+\Pi_y^2-2\Pi_xk_x^{t/b}-2\Pi_yk_y^{t/b}+|\boldsymbol{k}_{t/b}|^2
    \right)\\
    =&-\frac{\hbar^2}{2m^*}\left(
    \frac{1}{l_b^2}(2n+1)
    +\frac{1}{3}g^2
    \right)+\frac{\hbar^2}{m^*}\left(
    \Pi_xk_x^{t/b}+\Pi_yk_y^{t/b}
    \right)\,.
\end{split}
\end{equation}

Note that the ladder operators for $\boldsymbol{\Pi}$ and $\boldsymbol{\eta}$ are not the same. We have
\begin{equation}
    \begin{split}
        &\frac{\hbar^2}{m^*}\left(
    \Pi_xk_x^{t/b}+\Pi_yk_y^{t/b}
    \right)\\
    =&\frac{\hbar^2}{m^*}\left(
    \frac{\eta_y}{\ell^2}k_x^{t/b}-\frac{\eta_x}{\ell^2}k_y^{t/b}
    \right)\\
    =&\frac{\hbar^2}{m^*}\left(-i
    \frac{a-a^{\dagger}}{\sqrt{2}\ell}k_x^{t/b}-\frac{a+a^{\dagger}}{\sqrt{2}\ell}k_y^{t/b}
    \right)\\
    =&\frac{\hbar^2}{m^*}\frac{1}{\sqrt{2}\ell}\left((-k_y^{t/b}+ik_x^{t/b})a^{\dagger}+(-k_y^{t/b}-ik_x^{t/b})a
    \right)\,,
    \end{split}
\end{equation}
where the ladder operators do
\begin{equation}
\begin{split}
a^{\dagger}\left|\lambda,n,y\right\rangle &= \sqrt{n+1}\left|\lambda,n+1,y\right\rangle,\\
a\left|\lambda,n,y\right\rangle &= \sqrt{n}\left|\lambda,n-1,y\right\rangle.
\end{split}
\end{equation}

Using Fourier transform, according to
\begin{equation}
    \left|\lambda, n, y_0, j, k_2\right\rangle=\frac{1}{\sqrt{N}} \sum_u e^{i k_2(u p+j) \Delta}\left|\lambda, n, y_0+(u p+j) \Delta\right\rangle\,.
\end{equation}
1. The parabolic terms are unaffected since they are independent of $y$.\\
\noindent
2. The matrix element of the Moir\'e potential and tunneling terms (see Eq.\eqref{Vmn}) is modified as
\begin{equation}
\begin{split}
&\langle \lambda',n_2,y_0,j-m,k_2|V_{m\boldsymbol{g}_1+n\boldsymbol{g}_2}|\lambda,n_1,y_0,j,k_2\rangle\\
    =& e^{ik_2m\Delta}\langle n_2,y-m\Delta |V_{m\boldsymbol{g}_1+n\boldsymbol{g}_2}|n_1,y\rangle\,.
\end{split}
\end{equation}


\section{Chern-Simons Flux attachment}
\label{sec:appCSFluxAttachment}

We consider the two-dimensional system with $N_{uc} = N_1\,N_2$ unit cells, where $N_1$ and $N_2$ are, respectively, the number of unit cells extended along the directions of the primitive vectors $\bs{a}_1$ and $\bs{a}_2$. We denote the area of the system by $A_{system} = N_{uc}\,A_{uc}$, where $A_{uc} = |\bs{a}_1 \times \bs{a}_2|$ is the area of the unit cell.

Since the number of states per moir\'e band in each valley is given by $N_{uc}$, the band filling fraction per valley is denoted 
\begin{equation}
    \nu = \frac{N}{N_{uc}}
\,,    
\end{equation}
where $N$ is the number of particles per valley. 

The Chern-Simons coupling establishes a local relationship between the particle density $\rho$ and the Chern-Simons flux
\begin{equation}
\label{eq:local CS relation}
2\,\phi_0\,\rho = \nabla \times \bs{A} = B
\,.
\end{equation}
This relationship implies that each particle is attached to two flux quanta $2\,\phi_0\,$ of the Chern-Simons gauge field. (This can be generalized to the case where each particle is attached to an even number of flux quanta.)
Integrating Eq. \eqref{eq:local CS relation} over the total system of area, 
while considering a uniform CS field $B = $,
gives $2\,\phi_0\,N = B\,A_{u.c}\,N_{u.c}
$, which implies that the moir\'e band filling per valley is related to the Chern-Simons flux per unit cell
\begin{equation}
\label{eq: filling nu}
\nu = \frac{1}{2}\frac{\phi}{\phi_0}
=
\frac{1}{2}\frac{p}{q}
\,.
\end{equation}

We now relate the moir\'e filling $\nu$ to the filling of bands formed due to the interplay of the moir\'e potential and the CS field. For that, recall that a uniform field gives rise to Landau levels supporting $N_{LL} = B\,A_{system}/\phi_0$ states per LL, which can be re-expressed as 
$N_{LL} = B\,A_{system}/\phi_0
=
\frac{B\,A_{uc}}{\phi_0}\,N_{uc}
=
\frac{\phi}{\phi_0}\,N_{uc}
=
\frac{p}{q}N_{uc}
$.
Furthermore, we form the magnetic unit cell by enlarging the moir\'e unit cell by a factor of $q$, so that the number of states in the magnetic unit cell is $(q\,A_{uc})\,\rho_{LL} = (q\,A_{uc})\, \frac{1}{2\pi\ell^{2}_{B}} = (q\,A_{uc}) \frac{1}{2\pi\frac{\hbar}{eB}} = (q\,A_{uc})\, \frac{B}{\phi_0} = q\,\frac{\phi}{\phi_0} = p$\,. That is, each LL gives rise to $p$ ``subbands", each of which supports a number of state
\begin{equation}
\label{SM: eq: N sub}
N_{sub} = \frac{N_{LL}}{p} = \frac{N_{uc}}{q} 
\,.
\end{equation}
Combining Eqs. \eqref{eq: filling nu}
and \eqref{SM: eq: N sub} gives the number of occupied subbands, which denotes the composite fermion filling factor
\begin{equation}
\label{eq: CF filling}
\nu_{\text{\tiny{CF}}}
=
\frac{N}{N_{sub}} = q\,\frac{N}{N_{uc}} = q\,\nu = \frac{p}{2}
\,.
\end{equation}

\noindent
\textbf{(a) $p$ is an odd integer.} In this case, $\nu_{\text{\tiny{CF}}} = \frac{1}{2} + \mathbb{Z}$, corresponding to a half-filled CF band. This corresponds to even denominator filling fraction $\nu$, where CFs are gapless and form a Femi surface.
\\

\noindent
\textbf{(b) $p$ is an even integer.} We denote $p = 2\,m_1$ and write $q = 2m_2+1$, since $q$ must be odd since $(p,q)=1$. Then the filling fraction is $\nu = \frac{1}{2}\frac{p}{q} = \frac{m_1}{2m_2+1}$, which is an odd denominator filling fraction. From Eq. \eqref{eq: CF filling}, the number of filled composite fermion bands is $\nu_{\text{\tiny{CF}}} = m_1$. When the $m_1$-th and $(m_1+1)$-th CF bands are separated by a gap, an incompressible CF state arises, and a Chern number of this state can be computed at the corresponding filling fraction. 

\section{Quantum Phase Transitions}
\label{sec:moreQPT}

Due to the fractal nature of the moir\'e Hofstadter spectrum, we also observed quantum phase transitions at hole filling fractions $\nu_h = 1/5$ and $2/9$, i.e., flux $\phi/\phi_0 = 2/5$ and $4/9$, as shown in Fig.\,\ref{SM: fig: QPT 21} and Fig.\,\ref{SM: fig: QPT 389}. It is worth pointing out that the filling fraction follows a $\frac{p}{4p+1}$ sequence.

\begin{figure}[!htb]
    \centering
    \includegraphics[width=1\linewidth]{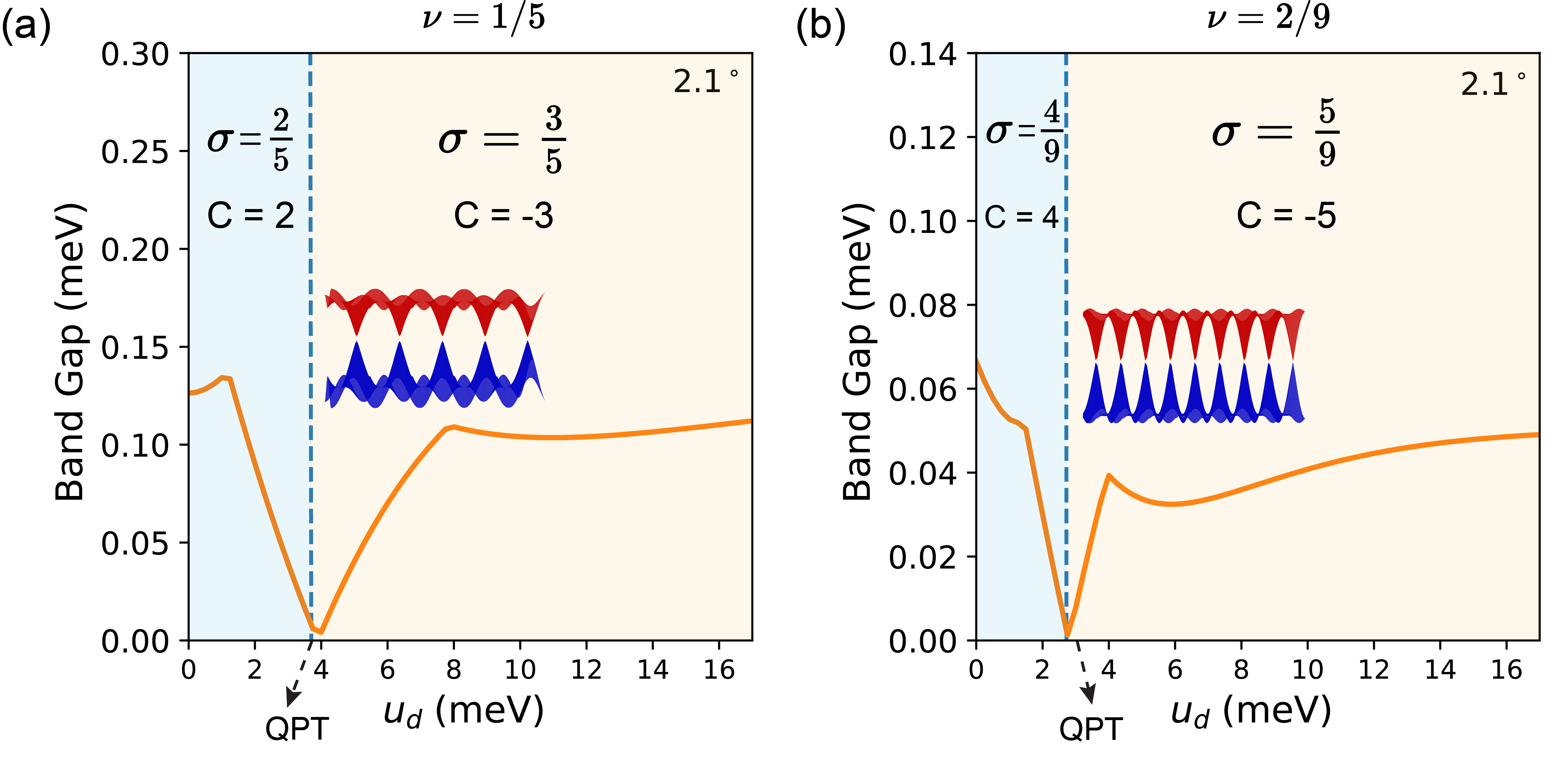}
    \caption{Composite fermion band gap at $2.1^{\circ}$. At filling (a) $\nu=\frac{1}{5}$, (b) $\nu=\frac{2}{9}$ (Fractal-FCIs), one quantum phase transitions occur, during which the occupied band (red) and the unoccupied band (blue) exchange Chern number through (a) five, (b) nine Dirac cones.}
    \label{SM: fig: QPT 21}
\end{figure}

\begin{figure}[!htb]
    \centering
    \includegraphics[width=1\linewidth]{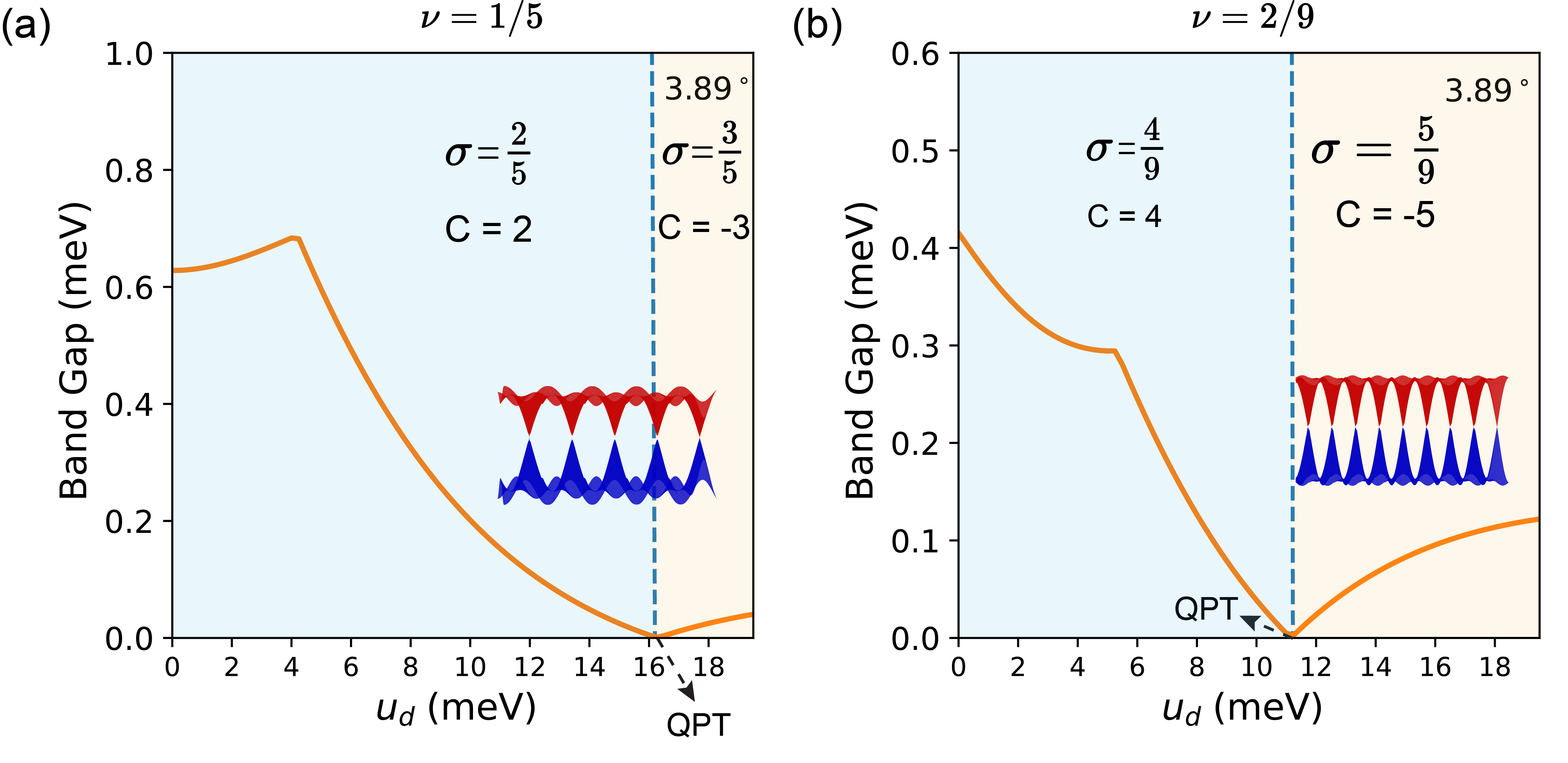}
    \caption{Composite fermion band gap at $3.89^{\circ}$. At filling (a) $\nu=\frac{1}{5}$, (b) $\nu=\frac{2}{9}$ (Fractal-FCIs), one quantum phase transitions occur, during which the occupied band (red) and the unoccupied band (blue) exchange Chern number through (a) five, (b) nine Dirac cones.}
    \label{SM: fig: QPT 389}
\end{figure}

\section{The second moir\'e band}
\label{sec:secondband}

In Fig.\,\ref{fig:second_moire} we present the Hofstadter spectrum for the second moir\'e band at $\theta\approx 2.1^\circ$, with the red curve tracking the Fermi level as the CS flux $\phi$ varies. 
Although the first and the second moir\'e bands appear similar in the zero-field limit, as shown in Fig.\ref{fig: moire bands}(a), the finite $b$ field spectrum of the second band, shown in Fig.\,\ref{fig:second_moire}, reveals striking differences compared to Fig.\ref{fig: Jain/Fractal 21}. Most notably, the composite fermion gaps, indicated by the jumps of Fermi level (red curve), are much smaller than those in the  first moir\'e band. This implies the topological orders from partially filling the second moir\'e band are more fragile, which is consistent with the fact all the salient FCI states in t-MoTe$_2$ observed in experiments are in the first moir\'e band. 

\begin{figure}[!htb]
    \centering
    \includegraphics[width=0.7\linewidth]{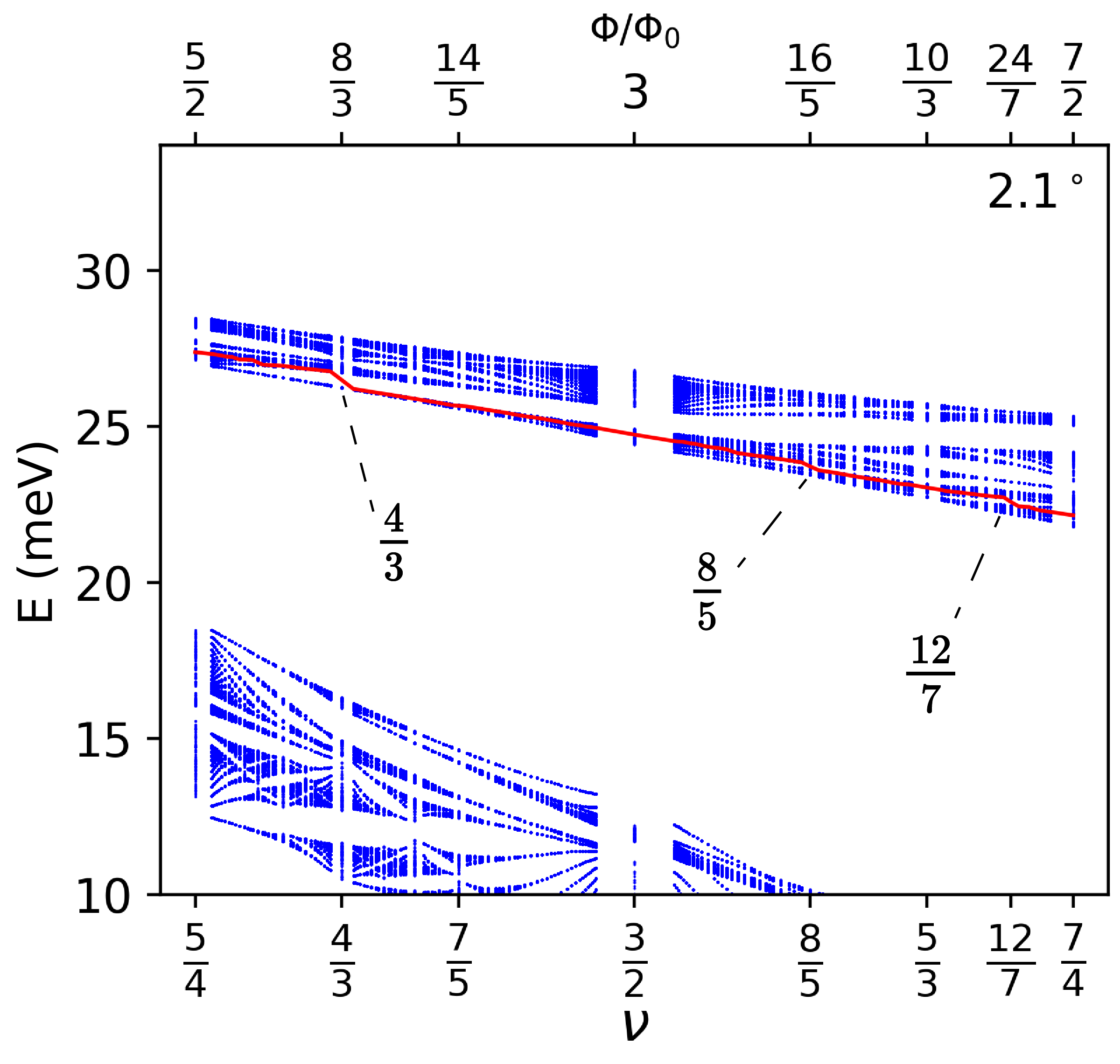}
    \caption{Hofstadter spectrum at $\theta \approx 2.1^{\circ}$ and $u_d=0$. The fermi surface is denoted as the red curve.}
    \label{fig:second_moire}
\end{figure}

\section{Magnetic Translation Symmetry}
As is well known, the presence of the magnetic field causes the Hamiltonian not to be invariant under regular translations. Let $W(\bs{a}) \equiv e^{i\bs{p}\cdot a/\hbar}$ be the generator of translations by a vector $\bs{a}$, such that 
$W(\bs{a})\,\bs{r}\,W^{-1}(\bs{a}) = \bs{r} + \bs{a}$\,. Then,
$W(\bs{a})\,\bs{\Pi}\,W^{-1}(\bs{a}) = \frac{1}{\hbar}(\bs{p} + e\bs{A}(\bs{r}+\bs{a})) = \frac{1}{\hbar}(\bs{p} + e\bs{A}(\bs{r}) + e\,\bs{\nabla}f(\bs{r}))$, where we have used the gauge equivalence $\bs{A}(\bs{r}+\bs{a}) = \bs{A}(\bs{r}) + \bs{\nabla}f(\bs{r})$ that is valid when the magnetic field is constant. The gradient of $f$ term can be ``removed", or gauged away, by a (unitary) gauge transformation
\begin{equation}
e^{-i\,\frac{e}{\hbar}f(\bs{r})}\,H[\bs{A}(\bs{r}) + \bs{\nabla}f]\,e^{i\,\frac{e}{\hbar}f(\bs{r})}  = H[\bs{A}(\bs{r})]   
\,.
\end{equation}
Then combining the regular translations and the gauge transformation gives
\begin{equation}
e^{-i\,\frac{e}{\hbar}f(\bs{r})}\,
W(\bs{a})\,
H[\bs{A}(\bs{r})]
\,
W^{-1}(\bs{a})
\,,
\,e^{i\,\frac{e}{\hbar}f(\bs{r})}  
= H[\bs{A}(\bs{r})]
\,.
\end{equation}
We then define the magnetic translation operator 
\begin{equation}
T_{\bs{a}} \equiv e^{i(\bs{p}\cdot\bs{a}/\hbar +\frac{e}{\hbar}f(\bs{r}))}    
\end{equation}

For concreteness, take $\bs{A}(\bs{r}) = \frac{\bs{B}\times\bs{r}}{2}$, such that the gauge transformation reads $f(\bs{r}) = \left( \frac{\bs{B}\times\bs{a}}{2} \right) \cdot \bs{r}$. Then, the magnetic translation operators result
\begin{equation}
T_{\bs{a}} 
=
e^{i \bs{a} \cdot \left( \Pi + \frac{1}{\ell^{2}_{B}} \hat{\bs{z}} \times \bs{r}\right)}
\,,
\end{equation}
satisfying the commutation relation
\begin{equation}
T_{\bs{a}} \, T_{\bs{a}'}
=
e^{i \frac{\hat{\bs{z}} (\bs{a} \times \bs{a}')}{\ell^{2}_{B}}}
\,
T_{\bs{a}'} \, T_{\bs{a}}
\,.
\end{equation}

The moir\'e unit cell is spanned by lattice vectors $\bs{a}_1$ and $\bs{a}_2$ forming a unit cell area $A_{uc} = \hat{z}\cdot(\bs{a}_1 \times \bs{a}_2)$. We will investigate the situation where the flux per unit cell is a rational number $p/q$ of the flux quantum,
\begin{equation}
\Phi = B\,A_{uc} = \frac{p}{q}\Phi_0 = \frac{p}{q}\,2\pi\ell^{2}_{B}
\,,
\end{equation}
implying
\begin{equation}
\frac{1}{\ell^{2}_{B}} = 2\pi \frac{p}{q} \frac{1}{A_{uc}}
\,.
\end{equation}

Thus, 
\begin{equation}
T_{\bs{a}_1} \, T_{\bs{a}_2}
=
e^{i \frac{\hat{\bs{z}} (\bs{a}_1 \times \bs{a}_2)}{\ell^{2}_{B}}}
\,
T_{\bs{a}_2} \, T_{\bs{a}_1}
=
e^{i 2\pi \frac{p}{q}}
\,T_{\bs{a}_2} \, T_{\bs{a}_1}
\,.
\end{equation}
\bibliography{bibliography}

\end{document}